\PassOptionsToPackage{pdfpagelabels=false}{hyperref}
\documentclass[useAMS,usenatbib]{mnras}

\usepackage[pdftex]{graphicx}
\usepackage[dvipsnames]{xcolor}
\usepackage{amsmath}  
\usepackage{amssymb}  
\usepackage{upgreek}  
\usepackage{afterpage}  
\usepackage{times}  
\usepackage{paralist}  
\usepackage{pbox}  
\usepackage{booktabs}  
\usepackage[flushleft]{threeparttable}  
\usepackage{array}
\usepackage{mathtools}  
\usepackage{graphicx}
\usepackage{subfigure}

\usepackage{grffile}

\newcommand{\Msun}{{\rm M}_{\odot}}
\newcommand{\Mtwo}{M_{\rm 200}}
\newcommand{\Mfive}{M_{\rm 500}}
\newcommand{\Rtwo}{R_{\rm 200}}
\newcommand{\Rfive}{R_{\rm 500}}

\title[Brightest cluster galaxies in IllustrisTNG]{The growth of brightest cluster galaxies in the TNG300 simulation: dissecting the contributions from mergers and \textit{in situ} star formation}

\author[D. Montenegro-Taborda et al.]
{
	\parbox{18cm}{
	  Daniel Montenegro-Taborda,$^{1}$\thanks{E-mail: d.montenegro@irya.unam.mx}
	  Vicente Rodriguez-Gomez,$^{1}$
	  Annalisa Pillepich,$^{2}$ \\
	  Vladimir Avila-Reese,$^{3}$
	  Laura V. Sales,$^{4}$
	  Aldo Rodr\'iguez-Puebla$^{3}$
    and Lars Hernquist$^{5}$
	}
	\vspace{0.3cm} \\ 
	$^{1}$ Instituto de Radioastronom\'ia y Astrof\'isica, Universidad Nacional Aut\'onoma de M\'exico, A.P. 72-3, 58089 Morelia, Mexico \\
    $^{2}$ Max-Planck-Institut f\"{u}r Astronomie, K\"{o}nigstuhl 17, D-69117 Heidelberg, Germany \\
	$^{3}$ Instituto de Astronom\'ia, Universidad Nacional Aut\'onoma de M\'exico, A.P. 70-264, 04510 CDMX, Mexico \\
    $^{4}$ Department of Physics and Astronomy, University of California, Riverside, 900 University Avenue, Riverside, CA 92521, USA \\
	$^{5}$ Harvard-Smithsonian Center for Astrophysics, 60 Garden Street, Cambridge, MA 02138, USA \\
}

\begin{document}
%
%
\maketitle
\begin{abstract}
We investigate the formation of brightest cluster galaxies (BCGs) in the TNG300 cosmological simulation of the IllustrisTNG project. Our cluster sample consists of 700 haloes with $\Mtwo \geq 5 \times 10^{13} \, \Msun$ at $z=0$, along with their progenitors at earlier epochs. This includes 280 systems with $\Mtwo \geq 10^{14} \, \Msun$ at $z=0$, as well as three haloes with $\Mtwo \geq 10^{15} \, \Msun$. We find that the stellar masses and star formation rates of our simulated BCGs are in good agreement with observations at $z \lesssim 0.4$, and that they have experienced, on average, $\sim$2 ($\sim$3) major mergers since $z=1$ ($z=2$). Separating the BCG from the intracluster light (ICL) by means of a fixed 30 kpc aperture, we find that the fraction of stellar mass contributed by \textit{ex situ} (i.e. accreted) stars at $z=0$ is approximately 70, 80, and 90 per cent for the BCG, BCG+ICL, and ICL, respectively. Tracking our simulated BCGs back in time using the merger trees, we find that they became dominated by \textit{ex situ} stars at $z \sim $1--2, and that half of the stars that are part of the BCG at $z=0$ formed early ($z \sim 3$) in other galaxies, but `assembled' onto the BCG until later times ($z \approx 0.8$ for the whole sample, $z \approx 0.5$ for BCGs in $\Mtwo \geq 5 \times 10^{14} \, \Msun$ haloes). Finally, we show that the stellar mass profiles of BCGs are often dominated by \textit{ex situ} stars at all radii, with stars from major mergers being found closer to the centre, while stars that were tidally stripped from other galaxies dominate the outer regions.

\end{abstract}

\begin{keywords} methods: numerical -- galaxies: clusters: general -- galaxies: formation -- galaxies: evolution -- cosmology: theory.

\end{keywords}

\section{Introduction}\label{sec:intro}
\renewcommand{\thefootnote}{\fnsymbol{footnote}}

\renewcommand{\thefootnote}{\arabic{footnote}}

The most luminous (or massive) galaxies in the Universe are the ones lying near the centres of rich galaxy clusters, the so-called brightest cluster galaxies (BCGs). BCGs are close to the minima of the potential wells of their host clusters, as evidenced by the typically small spatial offset between the BCG and the peak of the X-ray emission by hot gas in the cluster \citep[][]{jones1984}. Therefore, their evolution is believed to be closely linked to that of their host clusters. 

The properties of BCGs are often thought to be different from those of ``normal'' elliptical galaxies. Most BCGs present an extended low surface brightness envelope (cD morphology) and are above the average mass--size relation for non-BCG early-type galaxies \citep[e.g.,][]{Tonry1987,vonderLinden+2007,Zhao+2015,Kluge+2020}.  There are also several pieces of evidence indicating that the luminosity of BCGs is inconsistent with just statistical sampling of the cluster galaxy luminosity function: they are too bright \citep[e.g.,][]{TremaineRichstone1977,More+2012,Hearin+2013,Rodriguez-Puebla+2013}, pointing to peculiarities in their formation process.

The formation of BCGs is believed to be extremely hierarchical, at least since $z\sim 1$, growing mostly by mergers with other galaxies rather than by \textit{in situ} star formation. This is suggested by observations \citep[e.g.,][]{Aragon-Salamanca1998,lidman2012evidence,BurkeCollins2013,oliva2014galaxy} and is broadly consistent with structure formation within the now standard $\Lambda$ cold dark matter ($\Lambda$CDM) cosmological model. This has been shown both via semi-analytic and semi-empirical models \citep[e.g.][]{DeLucia2007, guo2011,contini2014formation,Shankar+2015} as well as via cosmological hydrodynamical simulations of galaxies \citep{bahe2017, Pillepich2018a, ragone2018bcg, Henden2020}, which also naturally explain the stellar mass growth of BCGs since $z\sim 1$ despite their relatively old stellar populations and passive star formation (SF) activity. Further evidence of merger-driven growth in BCGs comes from the observation of bright companions that are expected to merge within relatively short time-scales according to dynamical arguments, or from the observed spatial offset between the position of the BCG and the centre of the cluster as inferred from X-ray emission \citep[e.g.][]{lauer2014_BCG}.

Another important characteristic of BCGs is that they are generally surrounded by a very extended structure of diffuse light, the so-called intracluster light (ICL). The existence of the ICL was originally noted by \cite{zwicky1951_ICL} in the Coma cluster and has since been shown to be present in virtually all galaxy clusters observed with sufficiently deep imaging \citep[see][for a recent review]{montes2022Nature}. The ICL is often considered to be composed of stars orbiting `freely' in the gravitational potential of the cluster, not being bound to any individual galaxy. Observationally, it is difficult to separate the ICL from the stellar components of the BCG and other cluster galaxies, which has resulted in a diverse range of operational definitions for the ICL in the literature (e.g. \citealt{gonzalez2013galaxy,presotto2014,montes2018intracluster,zhang2019, kluge2021photometric}; Brough et al., in preparation). Irrespective of the exact definition, however, ICL stars are believed to originate from merger events and tidal stripping of satellite galaxies within the cluster. This scenario is supported by both theory \citep[e.g.][]{murante2007, Contini2018} and observations \citep[e.g.][]{montes2018intracluster, demaio2018}. Therefore, the ICL holds clues about the merging history, stellar mass assembly, and dark matter dynamics of the cluster.

While observational results support the idea that mergers are important for the mass growth of the BCG and ICL, it is difficult from observations alone to quantify this growth and the contribution from mergers, especially due to progenitor bias \citep[e.g.][]{Shankar+2015}. In fact, even in numerical simulations it is difficult to distinguish between the two main accretion channels, stripping and mergers, since this separation is sensitive to the definition of when the merger actually starts, as discussed in \cite{Contini2018}. Despite this caveat, numerical simulations -- especially large-volume cosmological simulations -- are required to study the formation of such rare objects in a cosmological context. For example, using the Millennium Simulation \citep{springel2005simulations} in combination with a semi-analytic model of galaxy formation, \cite{DeLucia2007} studied the hierarchical formation of simulated BCGs.  It should be noted, however, that the \cite{DeLucia2007} model does not explicitly take into account the formation of the ICL, although it is known that the growth of the BCG and ICL are closely related \citep[e.g.][]{Pillepich2018a}. According to their model, half of the stellar mass in BCGs formed rather early ($z \sim 4$--$5$), mostly in other galaxie
s, while the BCGs themselves `assembled' half of their final stellar mass until much later times ($z \sim 0.5$).

In recent years, hydrodynamic cosmological simulations have become increasingly successful at producing galaxy populations with properties that resemble those of real galaxies. Examples of such simulations include the Horizon-AGN \citep{dubois2014_Horizon-AGN}, Illustris \citep{genel2014_Illustris, vogelsberger2014_MNRAS, vogelsberger2014_Nature}, Magneticum \citep{hirschmann2014}, and EAGLE simulations \citep{schaye2015Eagle}. The main advantage of these simulations is that they naturally produce the diversity of environments, evolutionary paths, and merging histories that play a role in galaxy formation, which are difficult to sample in a statistically meaningful fashion using smaller-volume (non-cosmological) simulations.

For example, \cite{Rodriguez-Gomez2016} determined the origin of every stellar particle in the Illustris simulation, classifying them as \textit{in situ} (formed in the same galaxy where they are currently found) or \textit{ex situ} (formed in another galaxy and subsequently accreted via mergers or tidal stripping). In this context, they found that the \textit{ex situ} stellar mass fraction is a strong function of galaxy stellar mass, ranging from $\sim$10 per cent for Milky Way-sized galaxies to over 80 per cent for the most massive galaxies in the simulation ($M_{\ast} \sim 10^{12} \, \Msun$). They also showed explicitly that \textit{in situ} stars dominate the inner regions of most galaxies, while \textit{ex situ} stars are more common in the outskirts, and explored the \textit{transition radius}, defined as the galactocentric distance where \textit{in situ} and \textit{ex situ} stars become equally abundant, as a function of stellar mass and \textit{ex situ} fraction. A similar analysis was later carried out by \cite{dubois2016horizon} and \cite{qu2017} for the Horizon-AGN and EAGLE simulations, respectively, finding broadly consistent results. With similar scientific goals , \cite{remus2021accreted} studied the stellar mass profiles of $M_{\ast} = 10^{10-12} \, \Msun$ galaxies from the Magneticum simulation with (48 Mpc/$h$)$^3$ volume and found that $\sim$9 per cent of them are in fact dominated by \textit{ex situ} stars down to the galactic centre. However, the volumes of these simulations are not large enough to produce a statistically significant sample of haloes in the cluster range, which is necessary for a robust, quantitative study of BCGs and their surrounding ICL.

More recently, building upon the successes of the original Illustris simulation, the IllustrisTNG project \citep{Marinacci2018, Naiman2018first, Nelson2018, Nelson2019, Pillepich2018a, Pillepich2019, springel2018} introduced magnetohydrodynamic simulations of three cosmological volumes carried out with an updated galaxy formation model \citep{Weinberger2017, Pillepich2018}. The largest of these simulations, known as TNG300, evolves a cubical volume of $\sim$300 comoving Mpc per side, which makes it ideal for studying rare objects such as BCGs.

The stellar mass content of groups and clusters of galaxies in the TNG300 simulation has already been extensively analysed by \cite{Pillepich2018a}, who showed that their main properties are in broad agreement with observations, barring possibly a too steep relationship between BCG stellar mass and underlying total halo mass. They quantified that total halo mass is a very good predictor not only of the BCG and the ICL stellar mass, with small scatter, but also of the whole stellar mass profiles beyond the inner few kiloparsecs. They extended the predictions for the \textit{ex situ} stellar mass fraction of \cite{Rodriguez-Gomez2016} up to $10^{15} \, \Msun$ haloes, showing that their BCGs are made of \textit{ex situ} material by more than 60 per cent even within their innermost 10 kpc. And they advocated for a practical and arbitrary separation, if any, between BCG and ICL, given that their physical origin is the same: mergers and stellar mass accretion. In a more recent work, \cite{pulsoni2021} studied the assembly of early-type galaxies in the TNG100 simulation ($\sim$100 Mpc per side) of the IllustrisTNG project, focusing on the kinematics and composition of their stellar haloes. They grouped their galaxies' stellar mass profiles into different `classes', finding that most of the profiles are dominated by \textit{in situ} stars at small radii, but also finding that $\sim$8 per cent of their objects, typically the most massive, are dominated by \textit{ex situ} stars at all radii. Also using the TNG300 simulation,  \cite{sohn2022co-evolution} studied the scaling relations between the stellar velocity dispersion of the BCG and the velocity dispersion of the cluster, finding overall agreement with observations at $z \lesssim 1$. Finally and importantly, \cite{ardila2021} performed a consistent comparison of the mass and mass profiles of massive central galaxies (stellar mass $\gtrsim 10^{11.4} \, \Msun$) from deep Hyper Suprime-Cam observations and from TNG100 at $z \sim 0.4$, finding them consistent to the $\sim 0.12$ dex level out to 100 kpc clustercentric distances. 

In parallel, hydrodynamic \textit{zoom-in} simulations have also been used to study BCGs and their ICL, targeting smaller samples of massive galaxy clusters \citep[e.g.][]{cui2014, cui2022, bahe2017, ragone2018bcg, alonso2020intracluster, bassini2020, Henden2020, marini2022ML}. For example, \cite{ragone2018bcg} presented an analysis of the growth and evolution of BCGs simulated within the DIANOGA project using different apertures to measure their stellar mass content, which they found to be in agreement with observational constraints. They also found that the assembly redshift of BCGs decreases with increasing aperture size, a result consistent with an inside-out formation scenario.

In this work, we expand upon the analysis of \cite{Pillepich2018a} and study the formation and evolution of BCGs in the TNG300 simulation of the IllustrisTNG project, focusing on the stellar mass composition of the BCG, ICL, and BCG+ICL, as well as their merging histories. Going beyond the analysis of the same TNG300 objects presented by \cite{Pillepich2018a}, we compare the TNG300 outcome to the results of selected other hydrodynamical and semi-analytic models (as available), consider the redshift evolution and assembly of the BCG, ICL and BCG+ICL components, and also study the stellar mass profiles of the simulated BCGs decomposed by their \textit{in situ} and \textit{ex situ} stars. 

We have organized this paper as follows. The simulation used in this work is described in Section \ref{sec:methodology}, along with its merger trees and stellar assembly catalogues. In Section \ref{sec:general_properties} we present some general properties of our simulated BCGs and carry out some comparisons to observations. The redshift evolution of the stellar mass content and SF rate (SFR) of our simulated BCGs is studied in Section \ref{sec:merger_trees}. In Section \ref{sec:stellar_profiles} we analyse the stellar mass profiles of our BCGs. Finally, we discuss our results and present our conclusions in Section \ref{sec:discussion_and_conclusions}.

\section{Methodology}\label{sec:methodology}

\subsection{The IllustrisTNG simulation suite}\label{subsec:illustrisTNG}
We use data from \textit{The Next Generation Illustris Project}\footnote{\url{www.tng-project.org}} \citep[IllustrisTNG,][]{Marinacci2018, Naiman2018first, Nelson2018, Nelson2019, Pillepich2018a,Pillepich2019, springel2018}, a suite of magnetohydrodynamic cosmological simulations carried out with the moving-mesh code \textsc{arepo} \citep{Springel2010-arepo}, which model the formation and evolution of galaxies within the $\Lambda$CDM paradigm. The cosmological parameters used are consistent with recent Planck measurements \citep{ade2016planck} : $\Omega_\textrm{m}$ = 0.3089, $\Omega_{\Lambda}$ = 0.6911, $\Omega_\textrm{b}$ = 0.0486, $h$ = 0.6774, $\sigma_8$ = 0.8159, $n_\textrm{s}$ = 0.9667. The initial conditions are obtained at $z=127$ using the \textsc{N-GenIC} code \citep{springel2005simulations,springel2015n}, based on a linear-theory power spectrum. 

IllustrisTNG presents a new galaxy formation model \citep{Weinberger2017, Pillepich2018} that includes gas radiative cooling, SF and evolution, chemical enrichment, stellar feedback, and supermassive black hole (BH) growth and feedback. This new model improves upon that of the original Illustris simulation \citep{vogelsberger2013model, torrey2014model}  by producing simulated galaxies with more realistic stellar masses and sizes, as well as galaxy groups and clusters with gas fractions in better agreement with observations \citep[e.g.][]{Pop2022}.

The IllustrisTNG project features the TNG50, TNG100, and TNG300 simulations, which follow the coevolution of dark matter (DM), gas, stars, and supermassive BHs within a cubical volume of approximately 50, 100, and 300 Mpc per side, respectively. In this work we use the highest-resolution version of TNG300, which models 2500$^3$ DM particles and approximately 2500$^3$ baryonic resolution elements (gas cells or stellar particles), which have masses $m_{\rm DM} \approx 5.9 \times 10^7 \, \Msun$ and $m_{\rm b} \approx 1.1 \times 10^7 \, \Msun$, respectively. The spatial resolution of the simulation is set by the gravitational softening length of DM and stellar particles, which is equal to 1.4 kpc at $z < 1$ (and is fixed in comoving units at $z > 1$, which means that the resolution is better at earlier times), as well as by the size of the gas cells, which is adaptive, with smaller cells at higher densities \citep[e.g.][]{Pillepich2019}.

The simulation output is stored in 100 snapshots corresponding to redshifts between $z=20$ and $z=0$. For each snapshot, the identification of haloes and subhaloes proceeds in two stages. First, the friends-of-friends algorithm  \citep[FoF;][]{davis1985evolution} selects groups of particles that are separated by a linking length equal to or smaller than 0.2 times the mean interparticle separation. These FoF groups, which we refer to as \textit{haloes}, roughly correspond to virialized structures as predicted by the spherical collapse model. Afterwards, the \textsc{subfind} algorithm \citep{Springel2001-SUBFIND, dolag2009substructures} is used to identify gravitationally bound overdensities within each FoF group, which we refer to as \textit{subhaloes}. The central subhalo identified by \textsc{subfind} is a special type of object, since particles that are not gravitationally bound to satellites are usually associated to this `background subhalo', as long as they are gravitationally bound to it. Therefore, from a numerical standpoint, the central \textsc{subfind} object not only contains the BCG, but can also be considered to include the ICL component of a galaxy cluster.

Throughout this paper, $\Mtwo$ ($\Mfive$) denotes the mass measured within the radius that encloses an average overdensity of 200 (500) times the critical density of the Universe. We use $\Mtwo$ to define our cluster sample in Section \ref{subsec:clustersample}, and $\Mfive$ for some comparisons to other works in Section \ref{sec:general_properties}.

\begin{table*}
\begin{center}
\begin{threeparttable}
\begin{tabular}{ m{3.5cm}  m{6cm} m{6.5cm} }
\hline
Name & Definition & Mass measurement(s) \\ \hline
\\
Cluster of galaxies
&
Any halo (FoF group) with a spherical-overdensity mass $\Mtwo \geq 5 \times 10^{13} \, {\rm{M}}_\odot$ at $z=0$.
&
$\Mtwo$, $\Mfive$, and $M_{\ast, 500}$
\\
\\
BCG+ICL & 
The stellar mass in a cluster that is \textit{not} locked in satellites. This corresponds to the stellar mass that is gravitationally bound to the main central galaxy, as identified by \textsc{subfind}, and is found within $\Rfive$.
&
$M_{\rm \ast, BCG+ICL} \equiv M_{\ast}$(central galaxy, $r<\Rfive$)
\\
\\
BCG (brightest cluster galaxy)  & 
Stellar mass that is gravitationally bound to the main central subhalo (i.e. galaxy), measured within a fixed (3D) aperture of radius 30 kpc. (Note, however, that in Sections \ref{subsec:sfr_evolution} and \ref{subsec:sfr_evolution_2} we adopt an aperture of 50 kpc to measure the SFR and sSFR, while from Section \ref{subsec:Mass_vs_z} onwards we focus on the BCG+ICL as a single entity, rather than attempting to isolate the BCG component.)
& 
$M_{\rm \ast, BCG} \equiv M_{\ast}$ (central galaxy, $ r < 30 \, {\rm kpc}$)
\\
\\
ICL (intracluster light) & 
Stellar mass that is gravitationally bound to the main central subhalo (i.e. galaxy), measured within radii $ 30 \, {\rm{kpc}}<r<\Rfive$.
&
$M_{\rm \ast, ICL} \equiv M_{\ast}$ (central galaxy, $30 \, {\rm kpc} < r < \Rfive$)
\\
\\
Satellite galaxies
& 
The stellar mass in a cluster that is gravitationally bound to satellites, measured within $\Rfive$.
&
$M_{\rm \ast, sats} = M_{\ast, 500} - M_{\rm \ast, BCG+ICL}$ \\
\\

\hline
\end{tabular}
\caption{Definitions of the main objects and stellar components used throughout this work. Most of these definitions are based on the work of \protect\cite{Pillepich2018a}. Note that, in Section \ref{sec:merger_trees}, we also consider the stellar mass within twice the stellar half-mass radius, $M_{\ast, 2 r_{\rm half, \ast}}$, which can be considered to represent the BCG plus the \textit{inner} ICL and which corresponds to an aperture of roughly 100 kpc at $z = 0$ (see also Appendix \ref{sec: appendix}). Also note that $M_{\ast, 500}$ includes all the stellar mass within $\Rfive$, including satellites.}
\label{tab: Definition components}
\end{threeparttable}
\end{center}
\end{table*} 

\subsection{Merger trees and stellar assembly catalogues}\label{subsec:mergertrees}

In order to study the formation and merging histories of our simulated galaxies, we make use of the `baryonic' version of the \textsc{SubLink} merger trees \citep{Rodriguez-Gomez2015}, which track stellar particles and star-forming gas cells across different snapshots in order to associate each galaxy to its progenitors and its (unique) descendant from adjacent snapshots. The main progenitor of each object is defined not as the most massive one, but as the one with the `most massive history' behind it, following \cite{DeLucia2007}, which results in more stable mass histories. A merger is assumed to take place when two branches of the merger tree join, and the corresponding merger mass ratio is defined as $\mu = M_2 / M_1$, where $M_1$ and $M_2$ are the stellar masses of the primary and secondary progenitors measured at the time when the secondary reached its maximum stellar mass. Given $\mu$, we classify mergers as major ($\mu \geq 1/4$), minor ($1/10 \leq \mu < 1/4$), and very minor ($\mu < 1/10$).

We also make use of the stellar assembly catalogues described in \cite{Rodriguez-Gomez2016}, which provide information about the origin of each stellar particle in the simulation. In particular, we make use of the following labels:

\begin{enumerate}
    \item \textit{In situ/ex situ}: A stellar particle is classified as \textit{in situ} if the galaxy in which it formed is found along the main branch of the galaxy where it is currently found. Otherwise, the stellar particle is labelled as \textit{ex situ}.
    \item Accreted via mergers/stripped from surviving galaxies: \textit{ex situ} stellar particles can be further divided into those that were accreted during a merger event and those that were stripped from a galaxy that has not (yet) merged. 
\end{enumerate}

The catalogues also provide the stellar mass ratio $\mu$ of the corresponding merger event. These stellar particle attributes provide a detailed view of the stellar mass composition of each object in the simulation, which can be combined with particle positions (or any other particle-level data) to study how the various stellar components are spatially distributed.

\subsection{The cluster sample}\label{subsec:clustersample}
We consider all TNG300 haloes with masses $\Mtwo \geq 5 \times 10^{13} \, {\rm M}_{\odot}$ at $z=0$, which yields a total of 700 objects. For comparison, the TNG100 simulation (of $\sim$100 Mpc per side) contains only 41 systems with $\Mtwo \geq 5 \times 10^{13} \, {\rm M}_{\odot}$ at $z=0$, which illustrates the statistical power of TNG300 for studying such unique objects \citep[see, for example, fig. 2 from][]{Pillepich2018a}.

In order to reduce the dependence of some of our results on halo mass, we will often split our cluster sample into those with $\Mtwo < 10^{14} \, \Msun$ (420 objects) and those with $\Mtwo \geq 10^{14} \, \Msun$ (280 objects).

\subsection{Stellar components}\label{subsec:comps}
For each cluster, we distinguish between three different stellar components: the BCG, the ICL, and satellite galaxies, as defined and summarized in Table \ref{tab: Definition components}.

In general, accurately separating these components is a challenging task that is still a topic of active research, both theoretically \citep[e.g.,][]{contini2021_Review} and observationally \citep[e.g.,][]{kluge2021photometric}. For simplicity, we define these structures according to their halocentric distance and subhalo membership, i.e. gravitational binding energy. Importantly, we assume the BCG and ICL to be composed of stellar particles that are gravitationally bound to the main central galaxy, i.e. the galaxy that sits at the deepest point of the gravitational potential. On the other hand, we consider satellite galaxies to be composed of stars that are gravitationally bound to other subhaloes, as determined by the \textsc{subfind} algorithm.

We define the BCG mass as the stellar content of the main central subhalo at radii $r < 30$ kpc (though for some observational comparisons, we also use $r <50$ kpc or $<2r_{\rm half, \ast}$). Similarly, we consider the ICL to be composed of stars that are gravitationally bound to the main central subhalo but that are located at $30 \, \textrm{kpc} < r < \Rfive$. Finally, the stellar mass of satellite galaxies consists of the stellar content in subhaloes other than the central, and we only consider those that are part of the FoF halo and lie within $r < \Rfive$. These definitions are similar to those used in other works based on simulations \citep[e.g.][]{Pillepich2018a, Henden2020}. Differently than in \citealt{Pillepich2018a}, we report stellar masses and BCG properties as directly returned by the TNG300 simulation, without applying any correction to account for resolution effects \citep[see also][]{Engler+2021}.

We will also study the redshift evolution of our BCG sample in Section \ref{sec:merger_trees}, although in this context we do not attempt to distinguish the BCG from the ICL, and instead consider either the \textit{total} stellar mass of the central \textsc{subfind} object or its stellar mass within $2 \, r_{\rm half, \ast}$, an aperture that scales with the size of the BCG+ICL. Similarly, in Section \ref{sec:stellar_profiles} we quantify the stellar mass profiles of our simulated BCGs out to $50 \, r_{\rm half, \ast}$, which typically exceeds $\Rfive$.

All of our measurements are carried out within spherical (3D) apertures. We present a comparison of the stellar masses measured within different apertures in Appendix \ref{sec: appendix} (see also \citealt{Pillepich2018a}).

\begin{figure*}
  \centering
	\includegraphics[width=17.7cm]{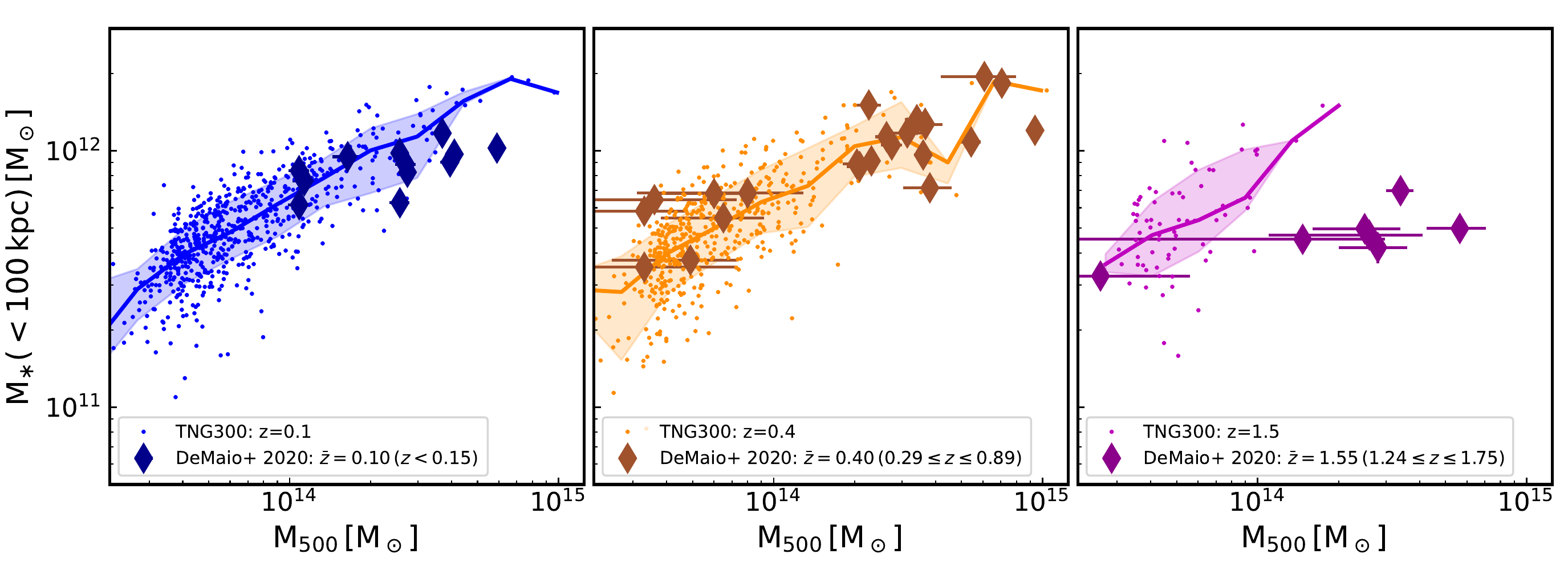}
	\caption{Stellar mass of the BCG+ICL within a fixed aperture of 100 kpc as a function of halo mass ($\Mfive$). The diamonds with error bars show observational estimates from \protect\cite{DeMaio2020} at $z < 0.15$ (left), $0.29 \leq z < 0.89$ (centre), and $1.24 \leq z \leq 1.75$ (right). These observational samples have mean redshifts of $\bar{z} \approx 0.1$, $\bar{z} \approx 0.4$, and $\bar{z} \approx 1.55$, respectively. The solid lines and shaded regions indicate the median and 16th to 84th percentile ranges of the simulated BCG+ICL systems at $z = 0.1$ (left), $z = 0.4$ (centre), and $z = 1.5$ (right), while the dots represent individual objects.}
	\label{fig:Mstellar-Mhalo in z}
\end{figure*}

\section{General properties of TNG300 BCGs}\label{sec:general_properties}

In this section, we compare the overall stellar mass content of BCGs in TNG300 to observational estimates at different redshifts, and also provide some global statistics about the merging history and stellar mass assembly of our simulated BCG sample.

\subsection{Stellar mass content in the BCG and ICL}\label{subsec: comparison to observations}

In Fig.~\ref{fig:Mstellar-Mhalo in z}, we plot the stellar masses of our simulated BCG+ICL systems, measured within a 3D aperture of 100 kpc, as a function of halo mass ($M_{\textrm{500}}$) at $z = 0.1$ (left), $z = 0.4$ (centre), and $z = 1.5$ (right). These measurements are compared to observational estimates by \cite{DeMaio2020}, within a projected aperture of 100 kpc, for different redshift intervals: $z < 0.15$ (left), $0.29 \leq z < 0.89$ (centre), and $1.24 \leq z \leq 1.75$ (right), which have mean redshifts comparable to those of our selected simulation snapshots ($\bar{z} \approx 0.1$, $\bar{z} \approx 0.4$, and $\bar{z} \approx 1.55$, respectively).

Overall, Fig.~\ref{fig:Mstellar-Mhalo in z} shows good agreement between the TNG300 simulation and the \cite{DeMaio2020} observations over a wide range of redshifts, with the possible exception of the highest redshift interval ($z \approx 1.5$, third panel). Note that, in the latter case, the TNG300 volume is not large enough to produce such massive clusters at high redshift, resulting in little overlap between the halo mass ranges of the simulated and observational samples at $z \approx 1.5$. Also note that here we compare 3D-aperture masses with 2D ones.

We will carry out more detailed comparisons to observations in upcoming work, taking into account possible selection effects in the cluster samples and performing measurements in `observer space' by means of forward-modelling of the simulation data \citep[e.g.][]{Rodriguez-Gomez2019}.

\begin{figure}
  \centering
	\includegraphics[width=7cm]{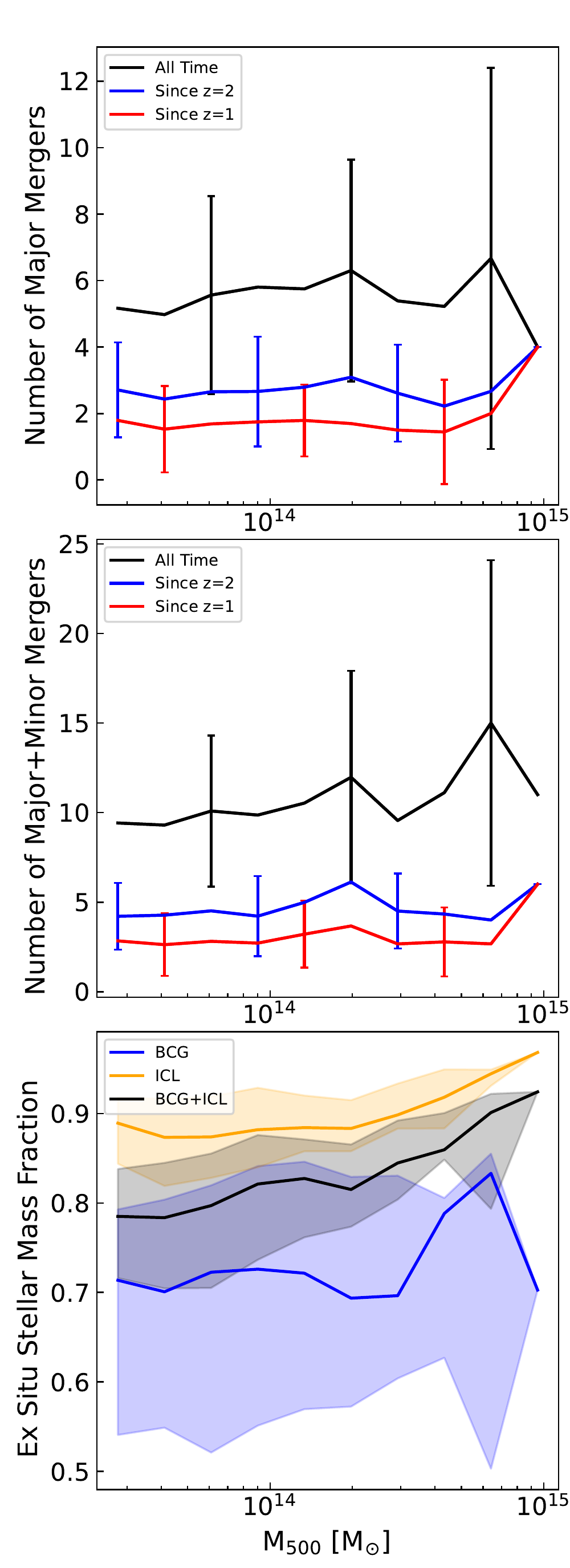}
	\caption{Merging history and stellar assembly statistics of the TNG300 simulated BCG sample as function of halo mass ($\Mfive$) at $z=0$. The upper two panels show the mean number of major (top) and major+minor (middle) mergers that our simulated systems have undergone since $z=1$ (red), $z=2$ (blue), and in total (black), with error bars indicating the cluster-to-cluster standard deviation. The bottom panel shows the median \textit{ex situ} stellar mass fraction of different components (BCG, ICL, and BCG+ICL, where the BCG consists of the stellar mass within 30 kpc, as defined in Table \ref{tab: Definition components}), with shaded regions indicating the corresponding 16th to 84th percentile ranges across the considered simulated systems.}
	\label{fig: number-mergers-and-ex-situ_star-fraction}
\end{figure}

\begin{figure*}
  \centering
	\subfigure[]{\includegraphics[width=8.5cm]{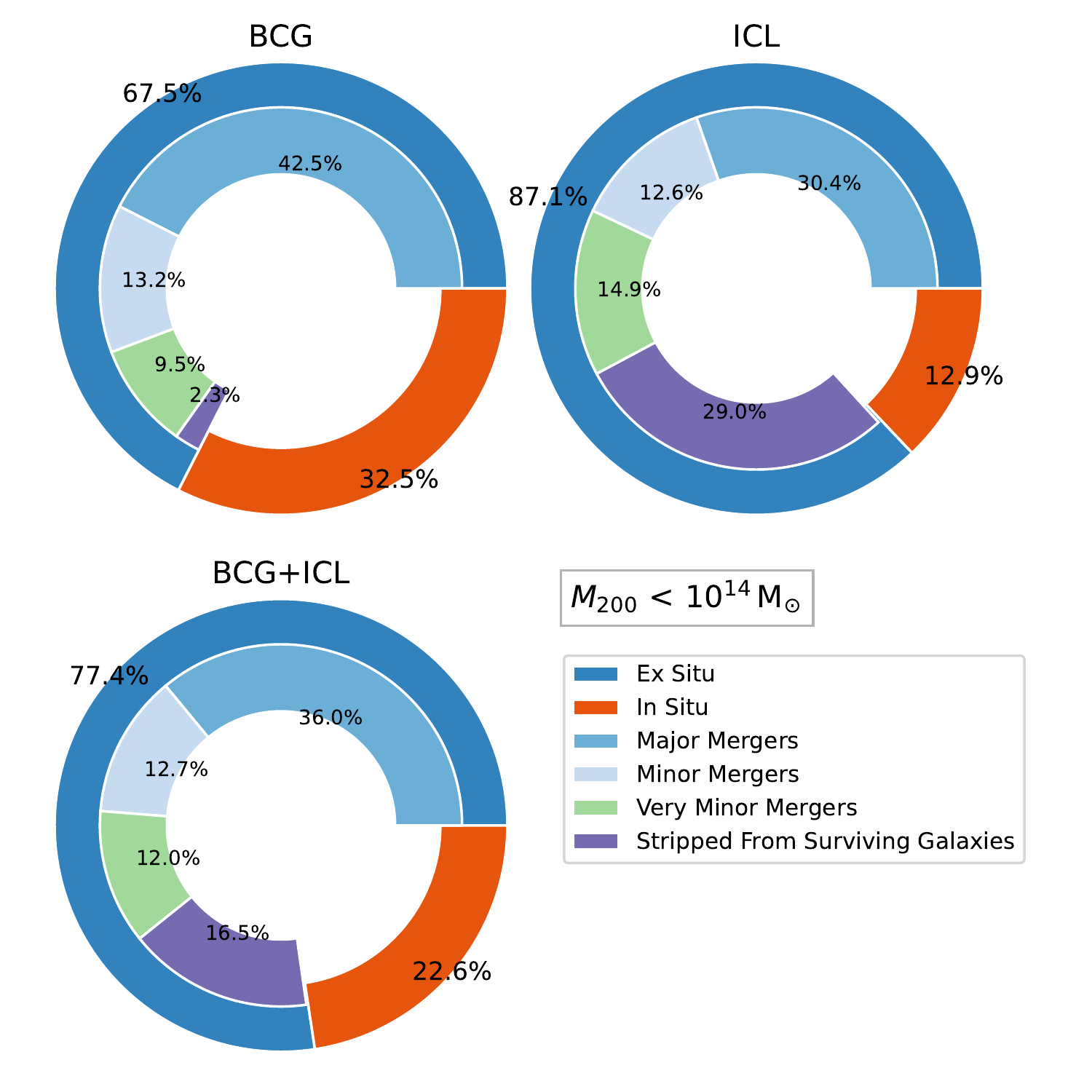}}
	\subfigure[]{\includegraphics[width=8.5cm]{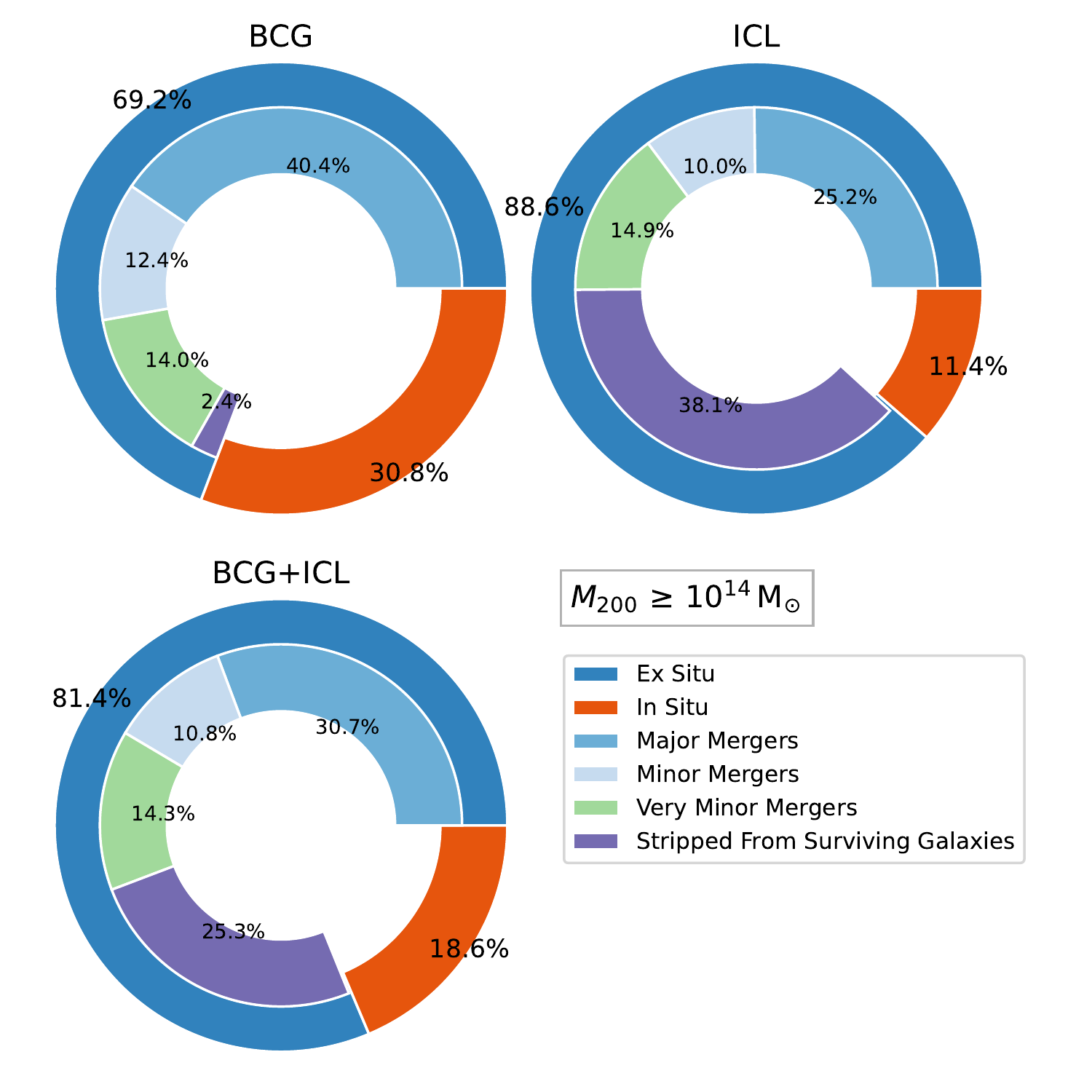}}
	\caption{Panel (a): Mean stellar mass `budget' of the BCG (top-left), ICL (top-right), and BCG+ICL (bottom-left) for all TNG300 clusters with $\Mtwo = 5 \times 10^{13} - 10^{14} \, \Msun$. Panel (b): Same as panel (a) but for $\Mtwo = 10^{14-15} \, \Msun$ clusters. In each pie chart, the outer level shows how the stellar mass splits into \textit{in situ} (orange) and \textit{ex situ} (blue) components, whereas the inner level separates the \textit{ex situ} stellar mass into the following components: stars accreted from (i) major mergers, (ii) minor mergers, (iii) very minor mergers, and (iv) stripped from surviving galaxies, as indicated in the legend. Note that the fractions in the second level are normalized with respect to the \textit{total} stellar mass (that is, including both the \textit{in situ} and \textit{ex situ} components). This figure shows that the contributions from \textit{in situ} stars and major mergers are higher in the BCG than in the ICL. On the other hand, the ICL contains a higher fraction of stars that were stripped from surviving galaxies (e.g. flybys or ongoing mergers). While the overall \textit{ex situ} stellar mass fractions in the BCG and ICL are similar between low and high mass clusters (left vs. right) ($\sim$70, $\sim$80, and $\sim$90 per cent for the BCG, BCG+ICL, and ICL, respectively), the stellar mass component in the ICL that comes from stars that were `stripped' from surviving satellites increases from $\sim$30 to $\sim$40 per cent for the more massive objects.}
	\label{fig:Pie-chart}
\end{figure*}


\subsection{Merging history}\label{subsec:merging_history}

Using the merger trees and stellar assembly catalogues described in Section \ref{subsec:mergertrees}, it is possible to study not only the growth and merging histories of individual BCGs, but also the origin of individual stellar particles, as we do throughout the rest of this paper. 

In Fig.~\ref{fig: number-mergers-and-ex-situ_star-fraction}, we present basic statistics predicted by the TNG300 simulation about the number of major and minor mergers that TNG300 BCGs have experienced, as well as their overall \textit{ex situ} stellar mass fraction at $z=0$.

The top panel of Fig.~\ref{fig: number-mergers-and-ex-situ_star-fraction} shows the mean number of major mergers (stellar mass ratio $\mu \geq 1/4$) as a function of halo mass ($\Mfive$), while the middle panel shows an analogous plot for major+minor mergers ($\mu \geq 1/10$). In each panel, different curves indicate the number of mergers that the BCGs have undergone in different time intervals: since $z=1$ (red), since $z=2$ (blue), and in total (black). Noting that there is a weak correlation between the number of mergers and halo mass, we find that TNG300 BCGs have experienced $\sim$2 major mergers since $z=1$, $\sim$3 since $z=2$, and $\sim$5 in their entire past history. The number of major+minor mergers is higher than the number of major mergers by a factor of $\sim$1.5--2.

The bottom panel of Fig.~\ref{fig: number-mergers-and-ex-situ_star-fraction} shows the \textit{ex situ} stellar mass fraction as a function of halo mass for the BCG (blue), ICL (yellow), and BCG+ICL (black). These fractions are approximately constant at 70, 80, and 90 per cent for the BCG, BCG+ICL, and ICL components of haloes with masses $\Mfive \lesssim 2 \times 10^{14} \, \Msun$, which represent the majority of our cluster sample. These values are consistent with those reported by \cite{Pillepich2018a} for TNG300 clusters. We also note that the \textit{ex situ} stellar mass fraction of the BCG displays a larger scatter than that of the ICL, indicating that there is more halo-to-halo variation in the stellar assembly of the BCG than in that of the ICL. In the next section, we explore the composition and origin of the \textit{ex situ} stellar mass in the BCG and ICL in more detail.

\subsection{Stellar mass composition at $z=0$}\label{subsec:stellar_mass_budget}

In Fig.~\ref{fig:Pie-chart}, we present the stellar mass `budget' predicted by the TNG300 simulation for different components of the central galaxy (BCG, ICL, and BCG+ICL) at $z=0$. In particular, we show the mean mass contributions from \textit{in situ} and \textit{ex situ} stars (first level of the pie charts), as well as from stars accreted via mergers of different stellar mass ratios and from stars that were tidally stripped from surviving galaxies (second level of the pie charts). To quantify possible dependencies of these results on total halo mass, we consider two halo mass ranges: $\Mtwo = 0.5$--$1 \times 10^{14} \, {\rm M}_{\odot}$ (left) and $\Mtwo \geq 10^{14} \, {\rm M}_{\odot}$ (right). 

These figures confirm that both the BCG and ICL are dominated by accreted stars, with \textit{ex situ} stellar mass fractions of approximately 70, 80, and 90 per cent for the BCG, BCG+ICL, and ICL, respectively, and displaying little variation between the two halo mass ranges considered. Additionally, the \textit{ex situ} component in Fig.~\ref{fig:Pie-chart} is separated according to whether the stars were accreted by means of mergers (major, minor, and very minor) or by stripping due to tidal forces from other galaxies, without there being a merger. For both halo mass ranges considered, we find that the contribution from major mergers is larger in the BCG ($\sim$40 per cent) than in the ICL ($\sim$25--30 per cent), while the fraction of stars that were tidally stripped from surviving galaxies is much higher in the ICL than in the BCG, representing 30--40 per cent of the ICL but only $\sim$2 per cent of the BCG. These findings are consistent with the theoretical expectation that the most massive satellite galaxies lose significant orbital angular momentum due to dynamical friction, reaching the centre of the cluster with possible events of stellar stripping followed by a major merger, whereas less massive satellite galaxies gradually lose material due to tidal forces as they orbit around the central object, depositing their loosely bound stars in the ICL.

We also note that there is a non-negligible amount of \textit{in situ} stellar material in the ICL, representing slightly more than 10 per cent of the total, which might seem counter-intuitive. We propose three possible explanations for this finding: (i) the aperture of 30 kpc that is used in this work is comparable to the effective radii of the most massive galaxies in the simulation \citep{genel2018size}, so that our definition of the ICL based on a fixed aperture might include part of the galaxy itself in the most extreme cases; (ii) major mergers can affect the dynamics of the stars in the galactic centre, changing their orbits and pushing some \textit{in situ} stars towards outer regions; (iii) accretion of cold gas in the ICL from satellite galaxies via tidal and ram-pressure stripping can lead to small outbreaks of SF \citep[e.g.][]{yun2019}. Ahvazi et al. (in prep.) will explore in more detail the physical processes responsible for \textit{in situ} SF in the ICL using the TNG50 simulation.


\section{Redshift evolution and stellar mass assembly according to TNG300}\label{sec:merger_trees}

How has the $z=0$ stellar mass budget of BCGs assembled across time? Throughout this section, we present results on the redshift evolution of the stellar mass, halo mass, and SFR of the TNG300 simulated BCG sample, as well as on the growth of the \textit{in situ} and \textit{ex situ} stellar components. We also revisit the concept of `assembly' and `formation' histories \citep{DeLucia2007}, which distinguish between the stellar mass assembled within the BCG itself (assembly history) and the stars belonging to the present-day BCG that had already formed by a given time (formation history).

For the purpose of studying the redshift evolution of BCGs and comparing our results to other works, we adopt slightly different definitions of stellar mass than those used in Section \ref{sec:general_properties}. In particular, we measure the SFR within an aperture of 50 kpc (instead of 30 kpc) when comparing to observations.
Similarly, when studying the redshift evolution of different stellar components of the BCGs and their ICL extensions, we will employ either the total stellar mass (i.e. the total gravitationally bound stellar mass as identified by \textsc{subfind}) or the stellar mass within twice the stellar half-mass radius, $r_{\rm half, \ast}$ (in order to use an aperture that `scales' with the physical size of the BCG). The latter definition can be considered to represent the BCG+ICL within a given aperture: the stellar mass within $2 r_{\rm half, \ast}$ is roughly comparable to the stellar mass within 100 kpc (see Appendix \ref{sec: appendix}), which is the definition of the BCG+ICL employed in the observations by \cite{DeMaio2020}. On the other hand, the total stellar mass of the \textsc{subfind} object can be assumed to correspond to the total stellar mass of the BCG+ICL extrapolated out to large radii, but without imposing an outer boundary. In addition, for the remainder of this paper, we will not attempt to separate the BCG from the ICL, and we will often refer to the BCG plus some ICL extension as simply the BCG.

Finally, in order to reduce the possible dependence of our results on halo mass, as well as to make closer comparisons to other works, throughout this section we will restrict our BCG selection to those with halo masses $\Mtwo \geq 10^{14} \, \Msun$ either at $z=0$ or at the given $z$, having verified that none of our results change significantly for the lower halo mass range $\Mtwo = 0.5$--$1 \times 10^{14} \, \Msun$. When comparing the assembly and formation histories of our BCGs to those of \cite{DeLucia2007} and \cite{Henden2020} in Section \ref{subsec:formation_vs_assembly_histories}, we will select haloes more massive than $\Mtwo = 5 \times 10^{14} \, \Msun$.

\subsection{SFRs of massive systems across cosmic epochs}
\label{subsec:sfr_evolution}

Although BCGs are traditionally believed to be `red and dead' objects, with essentially no \textit{in situ} SF \citep[e.g.,][]{Aragon-Salamanca1998,Donahue+2010,Fraser-McKelvie+2014}, recent observational studies have found that a non-negligible fraction of BCGs can have substantial amounts of SF activity, and that this fraction increases at higher redshifts \citep[e.g.,][]{mcdonald2016SFR, Bonaventura2017,Cooke+2019,Orellana-Gonzalez+2022}. 
Previous analyses have shown that the fraction of massive galaxies ($\gtrsim10^{11}\,\Msun$ in stars) that are quenched in the TNG300 simulation is in excellent agreement with data from COSMOS/UltraVISTA by \cite{Muzzin2013} at $z<1$ \citep{donnari2019} and with data from SDSS by \cite{Wetzel2012} at $z\sim0$ \citep{donnari2021b}, when accounting for observational definitions and selections. In this section, we expand on this topic within the IllustrisTNG framework and compare our results to additional observational results, particularly those focused on the highest mass end of observed objects.

\begin{figure}
  \centering
	\includegraphics[width=8.5cm]{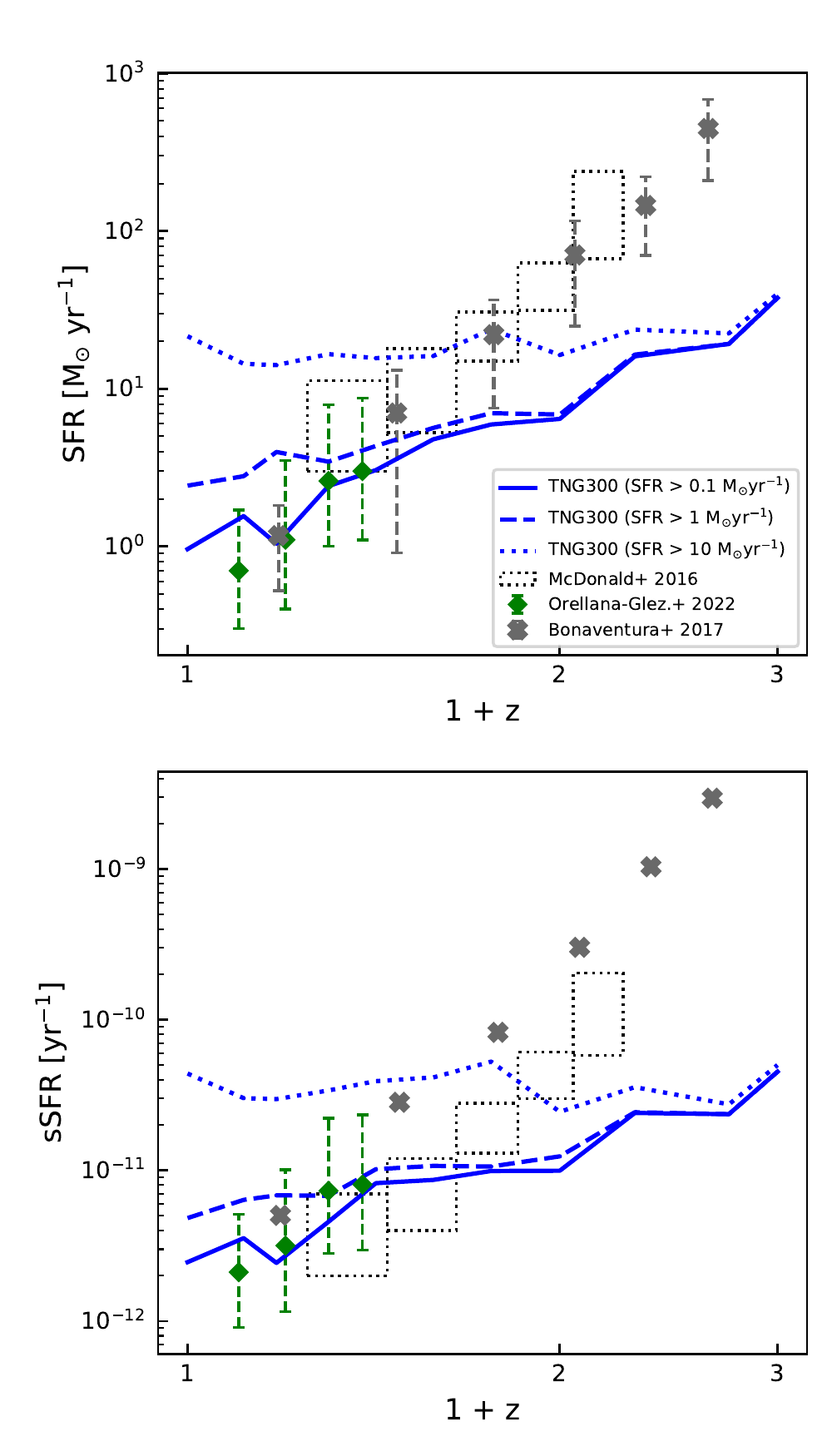}
	\caption{SFRs of BCGs across cosmic time according to the TNG300 simulation. In the top (bottom) panels, we show SFRs (specific SFRs). Median SFR (top) and specific SFR (bottom) as function of redshift, shown for TNG300 BCGs \textit{selected at each redshift} with $\Mtwo \geq 10^{14} \, \Msun$. The blue curves show measurements within 50 kpc from TNG300 including galaxies with three different SFR lower limits: 0.1 (solid), 1 (dashed), and 10 (dotted) $\Msun \, {\rm yr^{-1}}$. The black dotted rectangles, grey crosses and green diamonds represent observational estimates by \protect\cite{mcdonald2016SFR}, \protect\cite{Bonaventura2017} and \protect\cite{Orellana-Gonzalez+2022}, respectively, for SZ- or optically-selected clusters, and thus including a diversity of galaxy and halo mass distributions at different redshifts.
	\label{fig:SFR-vs-z_observation}}
\end{figure}

First, we extract results from TNG300 and compare them to inferences of SFR and specific SFR (sSFR) of BCGs from selected cluster samples at different redshifts, which are shown on Fig.~\ref{fig:SFR-vs-z_observation}. These observational studies refer mostly to clusters of masses $\Mtwo\gtrsim 10^{14}$ M$_\odot$ {\it at the redshift of observation}. Therefore, we measure the SFR and sSFR of BCGs at each redshift that reside in haloes of masses $\Mtwo \geq 10^{14}$ M$_\odot$. Typically, the observations we consider here cannot measure SFRs below some minimum value, and hence can only return upper limits in these cases. For example, \cite{Orellana-Gonzalez+2022} report SFR measurements down to $\sim$0.1 $\Msun \, {\rm yr}^{-1}$, while \cite{mcdonald2016SFR} quote a sensitivity better than $10 \, \Msun \, {\rm yr}^{-1}$ for $> 90\%$ of their sample. Considering this range of observational SFR thresholds, the solid, dashed, and dotted blue lines in Fig.~\ref{fig:SFR-vs-z_observation} show the population-wide median trends from TNG300 for BCGs with SFR $>$ 0.1, 1, and 10 $\Msun \, {\rm yr}^{-1}$, respectively, measured within an aperture of 50 kpc.

The dotted rectangles in Fig.~\ref{fig:SFR-vs-z_observation} show observational constraints by \cite{mcdonald2016SFR} for Sunyaev--Zel'dovich-selected (SZ), $X$-ray emitting clusters at $0.25 < z < 1.2$, obtained by combining observations in ultraviolet continuum, [O\textsc{ii}] line emission, and infrared (IR) continuum. Note that arithmetic averages rather than medians are reported in this particular case. The height of the boxes represents the combined statistical uncertainty in the mean and additional uncertainty due to non-detections. Their SFRs and stellar masses were estimated using a \cite{salpeter1955_IMF} initial mass function (IMF), so we have corrected them to the \citet[][]{chabrier2003galactic} IMF, for consistency with the IllustrisTNG model, which yields lower SFRs and stellar stellar masses by approximately 0.25 dex. The grey symbols with dashed error bars correspond to medians and 16th-84th percentiles by \cite{Bonaventura2017} for optically selected BCGs at $0 < z < 1.8$ using median flux stacks of IR-band spectral energy distributions (SEDs), which are fitted in order to estimate far IR (FIR) luminosities, $L_{\rm FIR}$, and from here, the SFRs. The SFRs and masses were also corrected from the IMF of Salpeter to that of Chabrier. Finally, the green diamonds with dashed error bars correspond to medians and 16th-84th percentiles by \citet[][]{Orellana-Gonzalez+2022} for a very large sample of BCGs at $0.05 < z < 0.42$ using SED fitting in optical and IR bands from SDSS and WISE, respectively.

We can see from Fig.~\ref{fig:SFR-vs-z_observation} that the SFRs and sSFRs of TNG300 BCGs are in good agreement with the low-redshift ($z \lesssim 0.4$) measurements by \citet[][]{Orellana-Gonzalez+2022} after selecting simulated galaxies with an SFR lower limit at $0.1 \, \Msun \, {\rm yr}^{-1}$ (solid blue line), roughly consistent with their ranges of measured SFRs. We note that, since the cluster sample of \citet[][]{Orellana-Gonzalez+2022} is based on optical selection of spectroscopically confirmed galaxies, we do not expect significant selection effects.

At higher redshifts ($z \gtrsim 1$), however, the predictions from TNG300 lie below the observational estimates by \cite{Bonaventura2017} and \cite{mcdonald2016SFR}, regardless of the SFR threshold. What could be the reason for this discrepancy, at least at face value? On the one hand, the TNG300 simulation might simply return too few star-forming massive galaxies at high redshift or galaxies with too low levels of SF at high redshift than in reality. On the other hand, however, the comparisons in Fig.~\ref{fig:SFR-vs-z_observation} are made at face value, i.e. without replicating the selection applied to the observational BCG sample. In fact, the \cite{Bonaventura2017} inferences may be affected by several effects: biases in the cluster selection method; contamination by late-stage mergers that are unresolved and seen as the BCG; sample incompleteness at $z>1.3$ (at higher $z$, only the BCGs that are brighter than the increasingly more luminous 24$\mu$m-inferred IR detection threshold are detected); and overestimation of the $L_{\rm FIR}$-inferred SFRs at $z<1$ in case of post-merger (post-starburst) BCGs, when the dust  remains heated by SF after 100 Myr. These different effects tend mainly to skew the obtained SFR distribution towards high values.
Regarding \cite{mcdonald2016SFR}, their results refer to arithmetic averages rather than medians. The latter are lower than the former when the distribution is skewed towards low values, as is the case in this sample, especially due to the significant number of non-detections in the used SFR tracers. Non-detections are due to the low observational sensitivity in these tracers, which imply a cut-off depth in SFR: only BCGs with SFRs above $\sim$10 M$_\odot$/yr are detected. Those without detection were assigned upper limit values. Therefore,  the \cite{mcdonald2016SFR} data shown in Fig.~\ref{fig:SFR-vs-z_observation} are expected to have lower values for the medians after properly modelling the upper limits, possibly being consistent with the results of the TNG300 simulation.

Finally, we note, although do not show, that the median SFRs of the TNG300 simulated BCGs (Fig.~\ref{fig:SFR-vs-z_observation}, top) are consistent with about half of the 11 BCGs at $0.2 \lesssim z \lesssim 0.5$ studied by \cite{fogarty2017} using data from the CLASH survey \citep{postman2012}. However, the BCGs from the other half of their sample display substantially higher SFR values, up to $\sim$100 $\Msun \, {\rm yr^{-1}}$. These high values are perhaps not surprising when considering that the BCG sample from \cite{fogarty2017} is, by construction, composed of star-forming systems with significant UV emission.

\subsection{The past SFRs of $z=0$ BCGs}
\label{subsec:sfr_evolution_2}
In what follows, we now measure the {\it  evolutionary tracks or histories} of SFR and sSFR for BCGs selected at $z=0$, following their progenitors along the main branch in the merger trees. 

\begin{figure}
  \centering
	\includegraphics[width=8.5cm]{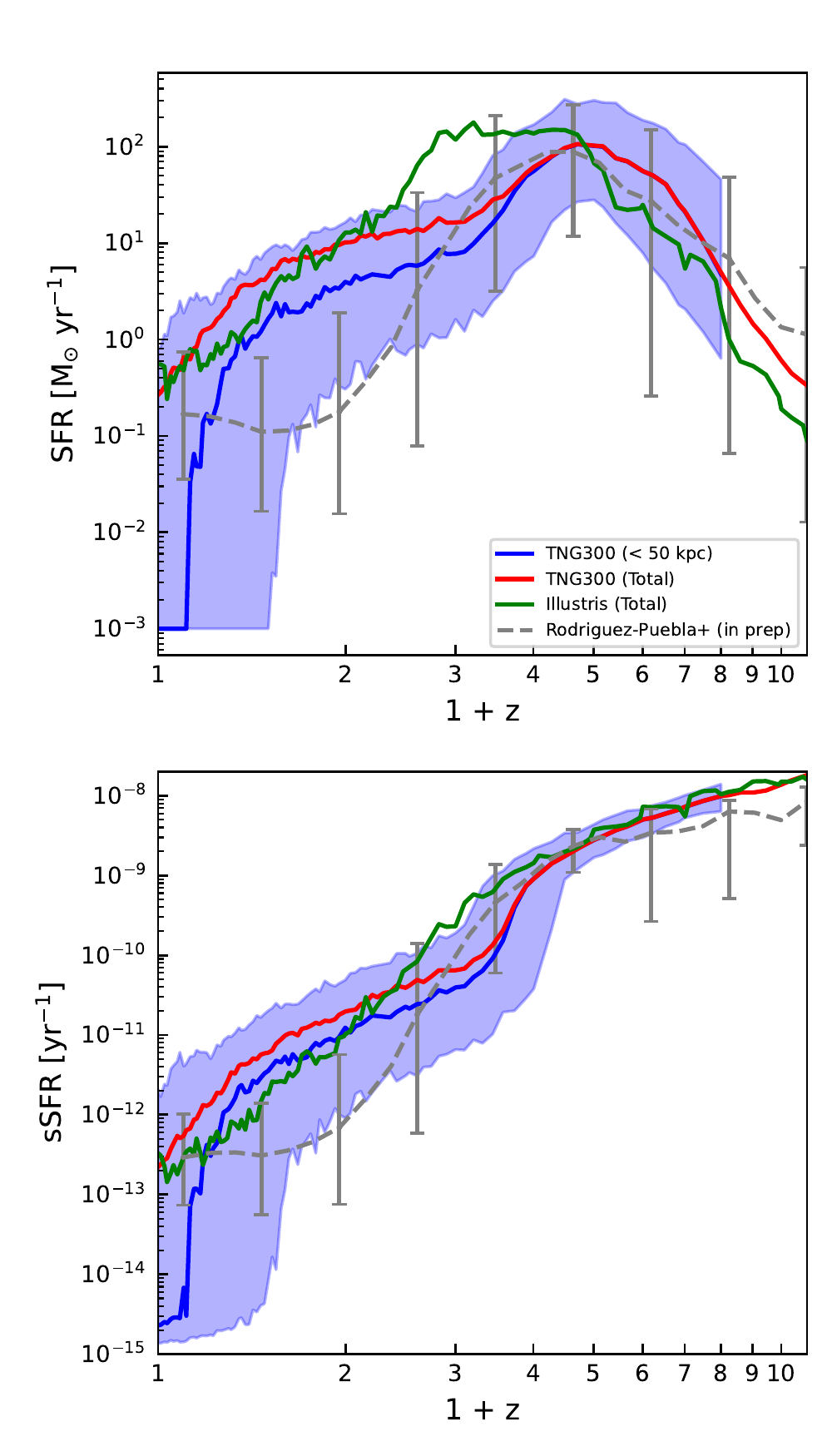}
	\caption{SF histories of BCGs across cosmic time according to the TNG300 simulation. In the top (bottom) panels, we show SFRs (specific SFRs).
	Median SFR (top) and specific SFR (bottom) as function of redshift, shown for the main progenitors of simulated BCGs selected at $z=0$ with $\Mtwo \geq 10^{14} \, \Msun$. The solid curves show measurements from TNG300 within 50 kpc (blue) and for the entire central galaxy (red), as well as for the entire central galaxy in Illustris original (green). The blue shaded regions indicate the 16th to 84th percentile range of the $r < 50$ kpc TNG300 measurements. The dashed grey line shows results from the semi-empirical model of Rodr\'iguez-Puebla et al. (in preparation). Null SFR values have been replaced with the minimum measurable values as discussed in \protect\cite{donnari2019}.
	\label{fig:SFR-vs-z_simulation}}
\end{figure}

In Fig.~\ref{fig:SFR-vs-z_simulation}, we show the corresponding median SFR (top panel) and sSFR (bottom panel) histories. The solid blue curves show measurements from TNG300 for an aperture of 50 kpc, while the shaded region indicates the corresponding 16th to 84th percentile range. For reference, we also show measurements for the entire \textsc{subfind} objects (without restricting the measurements to some aperture) for TNG300 (red) and Illustris original (green). The dashed grey line shows the median evolutionary tracks of SFR and sSFR of central galaxies in haloes more massive than $10^{14}$ M$_\odot$ at $z=0$ from the semi-empirical model by Rodríguez-Puebla et al. (in preparation). We note that, due to limitations in mass resolution, some simulated BCGs have ${\rm SFR} = 0$, which we have replaced with the smallest measurable values of $\rm SFR \approx 10^{-3} \, \Msun \, yr^{-1}$ for TNG300 and $\rm SFR \approx 10^{-4} \, \Msun \, yr^{-1}$ for Illustris original \citep[see][for a detailed discussion]{donnari2019, donnari2021a}, which can be regarded as upper limits of their `true' SFR values.

The semi-empirical results shown in Fig.~\ref{fig:SFR-vs-z_simulation} were obtained by means of a new and sophisticated evolutionary semi-empirical approach for the galaxy--halo connection (Rodriguez-Puebla et al., in preparation, which is an improved version of the model presented in \citealp{Rodriguez-Puebla+2017}). The model is based on the cosmological $N$-body simulation Bolshoi-Planck \citep[][]{Klypin2016} and on an extensive set of observational data from $z\sim 0$ to $z\sim 10$, including galaxy stellar mass functions dissected into star-forming and quiescent galaxies, $M_{\ast}-$SFR relations, the cosmic SF history, etc. The model includes parametric constraints for \textit{in situ} SFR and \textit{ex situ} stellar mass accretion, as well as ICL mass determination from mergers. By construction, the mock galaxy survey generated with the semi-empirical approach describes well the local and high-redshift galaxy population separated into star-forming/quiescent and central/satellite galaxies, including their spatial clustering; more details will be provided in Rodriguez-Puebla et al. (in preparation).

We can see from Fig.~\ref{fig:SFR-vs-z_simulation}, top panel, that the progenitors of BCGs rapidly increased their SFRs at very early epochs, reaching a maximum at $z \sim 3$--$4$, after which the SFRs decreased strongly. The bottom panel of Fig.~\ref{fig:SFR-vs-z_simulation} shows that the BCG progenitors at $z \gtrsim 3$ had sSFRs typical of star-forming galaxies at those epochs ($\gtrsim 10^{-10}$ yr$^{-1}$), and then, between $z \sim 3$ and $2$, their sSFRs strongly dropped. By $z \sim 1$, most of the BCG progenitors had already transitioned to a quiescent regime. These results are in good agreement with the semi-empirical inferences. Such an agreement is encouraging and may indicate that the early and rapid SFR decline in TNG300 BCGs due to the kinetic AGN feedback scheme may be realistic.

For similarly selected BCGs from the Illustris simulation, the maximum in SFR happens at $z \approx 2.5$, closer to the peak of the global SFR density \citep[e.g.][]{madau2014}. This suggests that the AGN feedback model of Illustris becomes effective at later times than that of IllustrisTNG -- in fact, the quenched fractions and red sequence at low-z of Illustris massive galaxies have been shown to be ruled out by observations \citep{vogelsberger2014_MNRAS, Nelson2018, donnari2019}, indicating a non-realistic implementation of supermassive BH feedback. At redshifts lower than $z \sim 2$, the semi-empirical inferences show a stronger decrease in SFR and sSFR than TNG300, at least down to $z \approx 0.6$, on average. If these inferences are correct, then the low but sustained SFR in the TNG300 BCGs may imply an inefficient AGN feedback mode for these galaxies at $z \lesssim 2$; at those epochs, the operating mode is already kinetic and the supermassive BH accretion rate is self-regulated.

In addition, we note that the sSFR decreases for smaller apertures (compare the blue and red lines in the lower panel of Fig.~\ref{fig:SFR-vs-z_simulation}), indicating that the inner regions of BCGs are more quenched than the outer ones. This is consistent with an inside-out quenching scenario \citep[e.g.][]{Nelson2019}, in which AGN feedback contributes first to the suppression of SF within the central regions of the BCGs. The above will be evident in the next subsections: such an inside-out quenching scenario postulated by the IllustrisTNG simulations has been shown to be supported by 3D-HST observations of $M_{\ast} \sim 10^{11}\,\Msun$ galaxies at $z\sim1$ \citep[e.g.][]{nelson2021}. At the same time, inside-out quenching makes comparisons of quenched fractions at high redshift extremely susceptible to where, within a galaxy, SF observations are sensitive to \citep{donnari2021b}.


\subsection{Mass growth of different cluster components}
\label{subsec:Mass_vs_z}

\begin{figure}
  \centering
	\includegraphics[width=8.1cm]{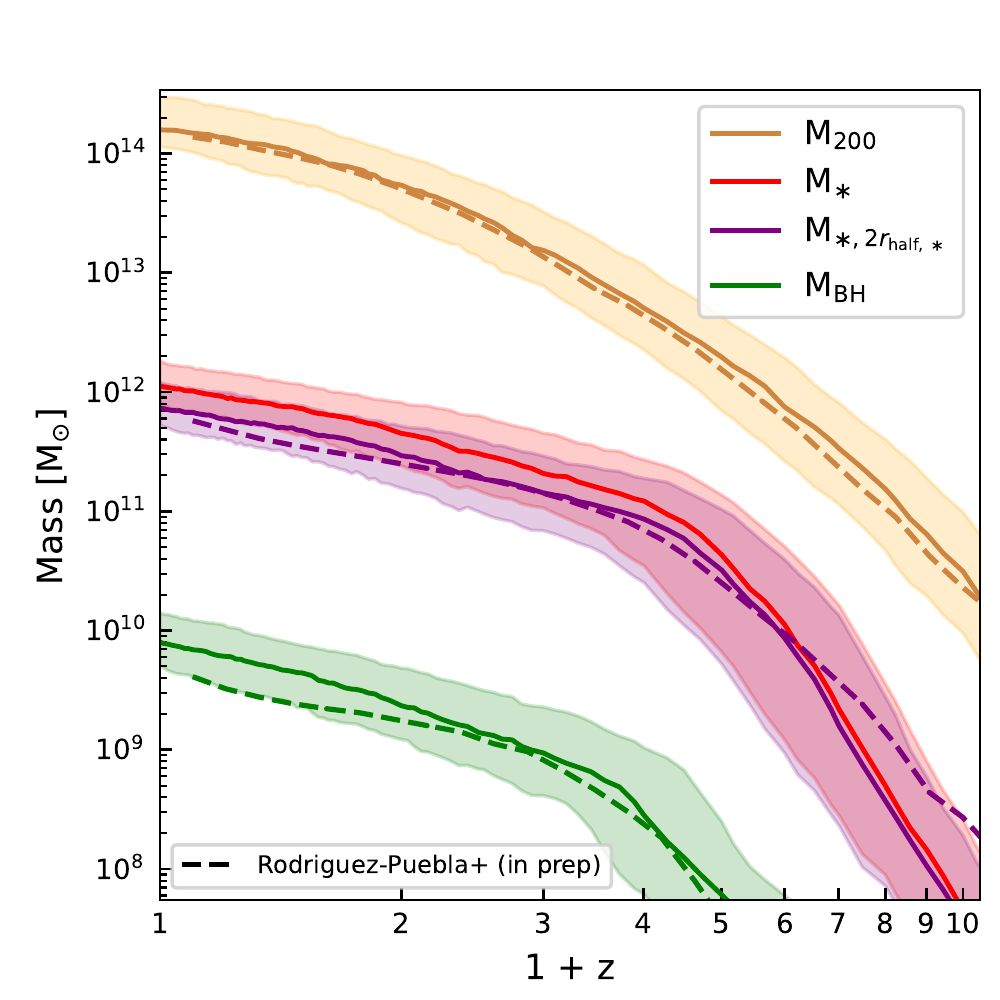}
	\caption{Median mass histories of BCGs in the TNG300 simulation selected at $z=0$ with $\Mtwo \geq 10^{14} \, \Msun$ and tracked back in time along the main branch of their merger trees. The solid red line shows the evolution of the total stellar mass that is gravitationally bound to the central main galaxy, while the solid purple line shows the stellar mass of the central main galaxy within $2 r_{\rm half, \ast}$. The solid orange line shows the mass evolution their host haloes' total mass, quantified by $\Mtwo$, while the solid green line shows the mass evolution of the supermassive BH hosted at the centre of the BCG. The shaded regions indicate the corresponding 16th to 84th percentile ranges, i.e. the galaxy-to-galaxy variation. The dashed lines represent the median mass histories for $\Mtwo$, $M_{\ast, 2 r_{\rm half, \ast}}$, and $M_{\rm BH}$ according to the semi-empirical model of Rodr\'iguez-Puebla et al. (in preparation). Note that, unlike the halo mass history, the shape of the stellar mass history displays an inflection at $z \sim 3$--$4$, such that the stellar mass growth rate slows down after this transition epoch.}
	\label{fig:mass-vs-z}
\end{figure}

In order to quantify the mass growth of the BCGs, their host haloes, and the supermassive BHs at their centres, we again make use of the TNG300 merger trees to track the main progenitors of BCGs selected at $z=0$ back in time. Fig.~\ref{fig:mass-vs-z} shows the median redshift evolution of various components for TNG300 haloes selected at $z=0$ with masses $\Mtwo \geq 10^{14} \, \Msun$. As anticipated, here we use different operational definitions of galaxy stellar mass than those of Table~\ref{tab: Definition components}. The different solid lines represent the total halo mass ($\Mtwo$, orange), the total stellar mass of the main central galaxy ($M_{\ast}$, red), the stellar mass of the main central galaxy within twice the stellar half-mass radius ($M_{\ast, 2 r_{\rm half, \ast}}$, purple), and the mass of the BH at the centre of the BCG ($M_{\rm BH}$, green). Note that the stellar mass estimates of Fig.~\ref{fig:mass-vs-z} include the smooth stellar components of both the BCG and ICL, in the nomenclature of the previous section, but within different effective apertures. The shaded regions indicate the corresponding 16th to 84th percentile ranges, while the dashed lines show predictions from the semi-empirical model by Rodriguez-Puebla et al. (in preparation).

It is evident from Fig.~\ref{fig:mass-vs-z} that the median masses of all the components grow monotonically with time. However, TNG300 predicts that the stellar mass growth shows a bend at $z \sim 3$--$4$, growing faster than the underlying DM halo mass at earlier times and slower at later times. This is indicative of much higher sSFRs at earlier times, consistent with Fig.~\ref{fig:SFR-vs-z_simulation}. The sSFR slows down at $z \sim 3$: in the case of TNG300, this has been shown to be caused by the ejection of the innermost gas and by the heating of the gas in the halo atmosphere, as a result of high-velocity gas outflows triggered by the kinetic BH feedback \citep[e.g.][]{Weinberger2017, Nelson2019, Zinger2020}.
The importance of the latter is supported by several observational studies, which show that BCGs have a high fraction of AGNs, mainly radio-loud, this fraction being higher in the past  \citep[e.g.,][]{Best+2007, Croft+2007, oliva2014galaxy, Yuan+2016}.


The evolution of the BH mass can provide further support to the possible causes of the decline in the SFRs of the BCGs. It can be seen from Fig.~\ref{fig:mass-vs-z} that the BHs grow in TNG300 rapidly in mass until $z\sim 3$, after which their growth slows down significantly. This corresponds to when the kinetic mode of AGN feedback kicks in at $z \sim 3$ \citep[see fig. 6 of][]{Weinberger2017}, which is characterized by lower accretion rates and more effective heating of the the gas surrounding the BCG, thus suppressing the SFR.

As seen in Fig.~\ref{fig:mass-vs-z}, the median mass histories of the BCGs, their cluster haloes, and their central BHs in TNG300 are in agreement with the semi-empirical inferences by Rodr\'iguez-Puebla et al. (in preparation), which by construction agree with a large body of observations at different redshifts, including SFRs, quasar luminosity functions, and BH mass correlations with stellar mass.


\begin{figure}
  \centering
    \includegraphics[width=8.5cm]{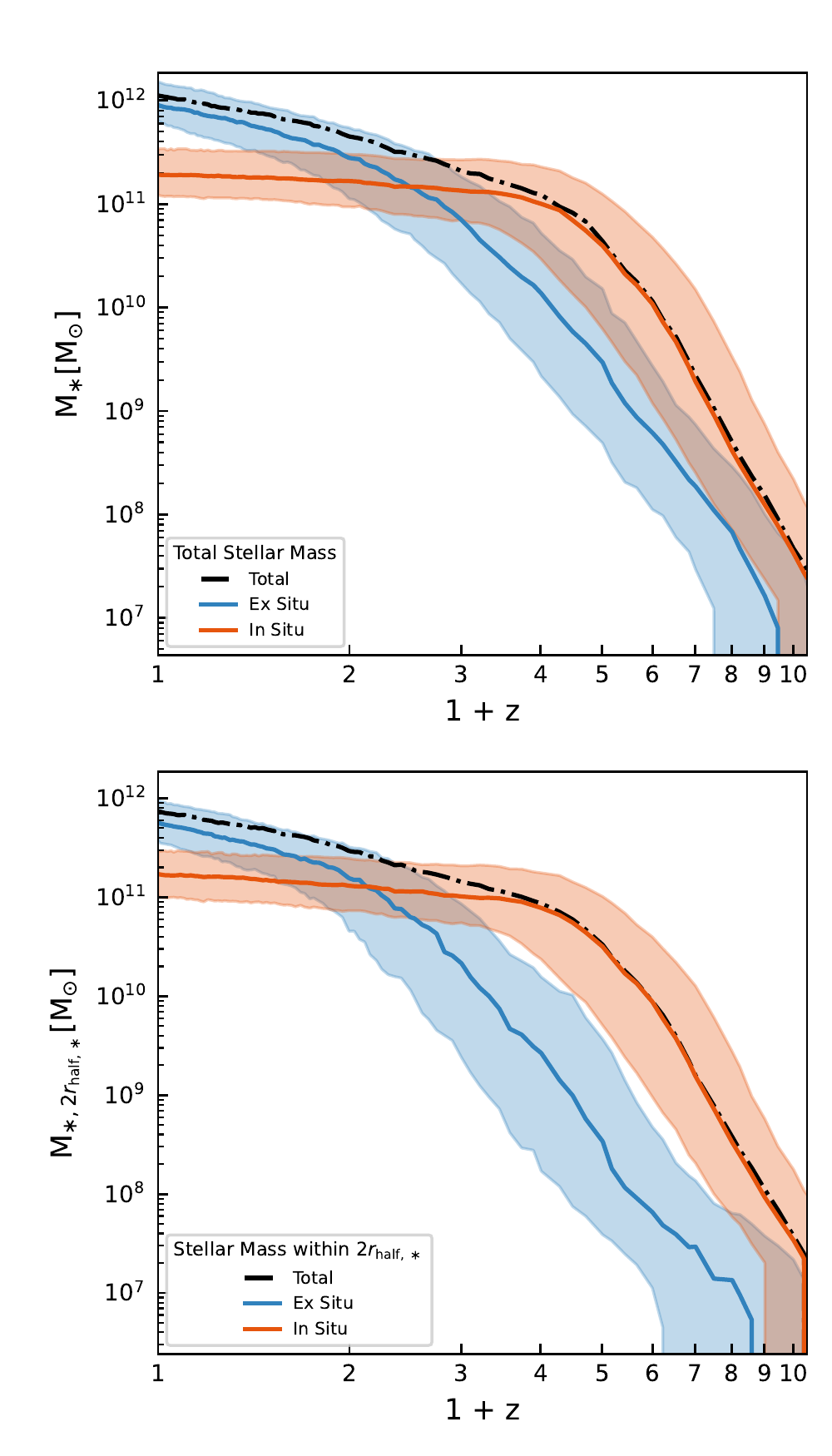}
    \caption{Redshift evolution of the total stellar mass (including the entire ICL; upper panel) and of the stellar mass within $2r_{\rm{half, \ast}}$ (including part of the ICL; lower panel) of TNG300 BCGs selected at $z=0$ with halo masses $\Mtwo \geq 10^{14} \, \Msun$. The orange and blue solid lines show the \textit{in situ} and \textit{ex situ} stellar components, while the black dot-dashed line shows the sum of both. The shaded regions indicate the 16th to 84th percentile ranges. Note that the stellar mass growth due to \textit{in situ} SF (orange) slows down significantly at $z \sim 3$--$4$, explaining the bend in the stellar mass history seen in Fig.~\ref{fig:mass-vs-z} within the same redshift range. On the other hand, the \textit{ex situ} stellar mass (blue) increases steadily with time, eventually surpassing the \textit{in situ} component. As a consequence, according to TNG300, the stellar mass growth of BCGs undergoes a transition from \textit{in situ}-dominated to \textit{ex situ}-dominated at $z \sim $1--2. }
	\label{fig:mstar(in-ex)situ-vs-z}
\end{figure}

\subsection{\textit{In situ} and \textit{ex situ} stellar mass growth}
\label{subsec:stellar_mass_growth}

The slowing down of the growth of the BCG stellar mass at $z \sim 2$--$3$ in TNG300 is also connected to the different time evolution of the \textit{in situ} and \textit{ex situ} stellar components. To illustrate this, the upper panel of Fig.~\ref{fig:mstar(in-ex)situ-vs-z} shows the median redshift evolution of the total stellar mass of TNG300 BCGs (black dot-dashed line, identical to red curve of Fig.~\ref{fig:mass-vs-z}), as well as that of its \textit{in situ} (orange) and \textit{ex situ} (blue) stellar components, where the shaded regions indicate the 16th to 84th percentile ranges. The lower panel is analogous to the upper one but for stellar masses measured within $2r_{\rm half, \ast}$. In both panels, the BCGs are selected at $z=0$ as those with halo masses $\Mtwo \geq 10^{14} \, \Msun$, and are tracked back in time using the merger trees.

As previously observed in Fig.~\ref{fig:mass-vs-z}, the shape of the stellar mass history exhibits a change in slope around $z \sim 3$--$4$, and the same bend is observed in Fig.~\ref{fig:mstar(in-ex)situ-vs-z} for the \textit{in situ} stellar component alone. Fig.~\ref{fig:mstar(in-ex)situ-vs-z} reveals that the growth of the \textit{ex situ} stellar mass does not present a similar change in slope, and instead increases steadily with time. This indicates that BCGs are continually accumulating stars that were formed in galaxies outside of their main branch in the merger trees, even if they have a negligible impact on stellar mass growth at early times. On the other hand, the \textit{in situ} stellar mass stops growing almost completely at $z\sim 3$--$4$, reflecting a drop in the SFR (quenching) as a result of AGN feedback.

At later times, $z \sim 1$--$2$, the cumulative contribution from \textit{ex situ} stars becomes larger than that of \textit{in situ} stars, resulting in a transition epoch between these two modes of stellar mass growth. As expected, the transition occurs slightly earlier when the entire ICL component is taken into account. By $z=0$, \textit{ex situ} stars represent the vast majority of the stellar content in the BCG+ICL, in agreement with the \textit{ex situ} stellar mass fractions presented in Sections \ref{subsec:merging_history} and \ref{subsec:stellar_mass_budget}.

These findings are broadly consistent with the so-called `two-phase' formation scenario of massive galaxies \citep[e.g.][]{Oser2010, Rodriguez-Gomez2016}: stellar mass growth at early times consists almost entirely of \textit{in situ} SF, but eventually transitions to a regime dominated by \textit{ex situ} stellar accretion, mostly in the form of gas-poor (`dry') mergers, which in turn have an impact on galaxy morphology and ultimately lead to the formation of massive early-type galaxies \citep{Rodriguez-Gomez2017}.


\begin{figure*}
  \centering
	\includegraphics[width=15cm]{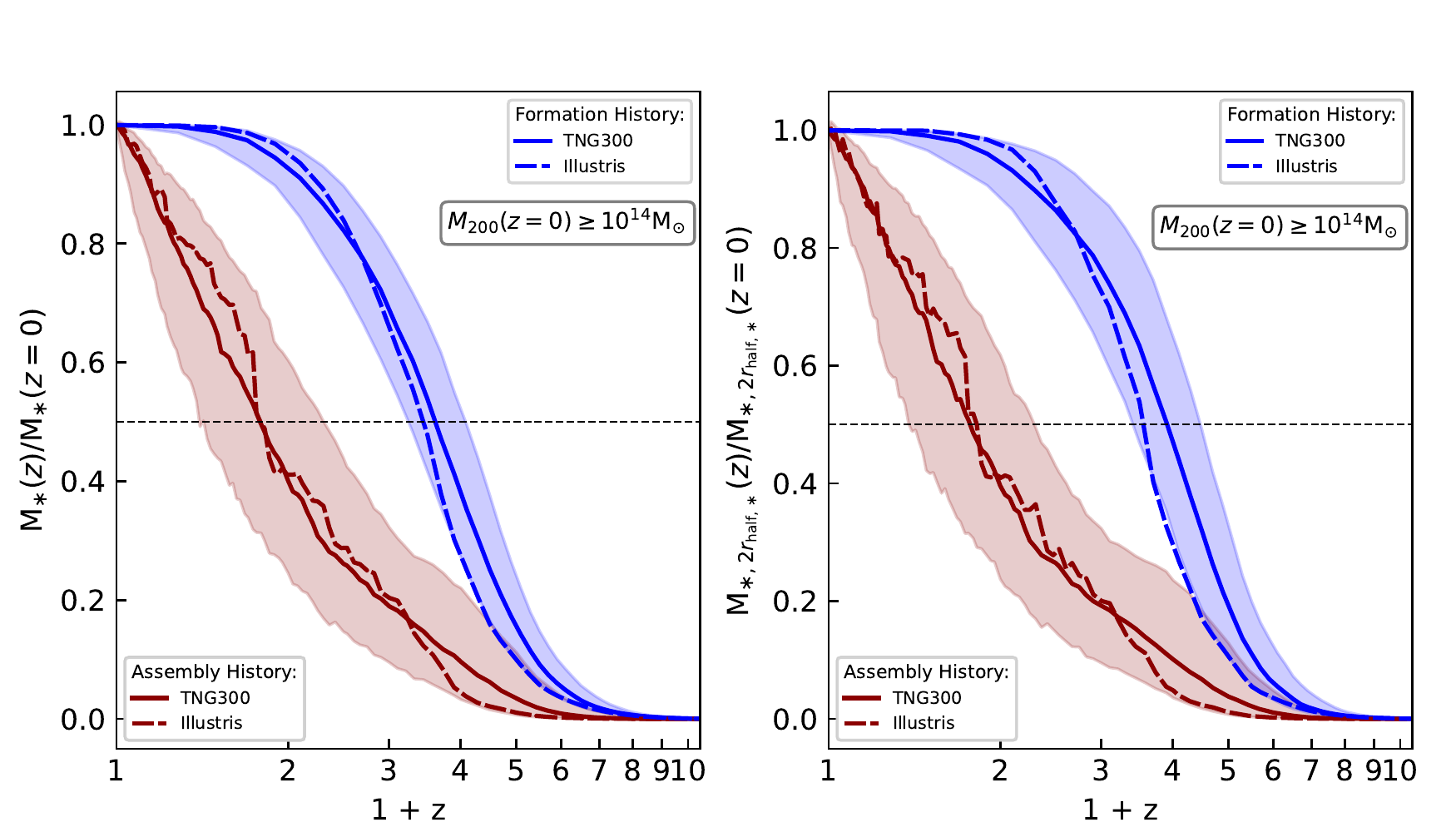}
	\includegraphics[width=15cm]{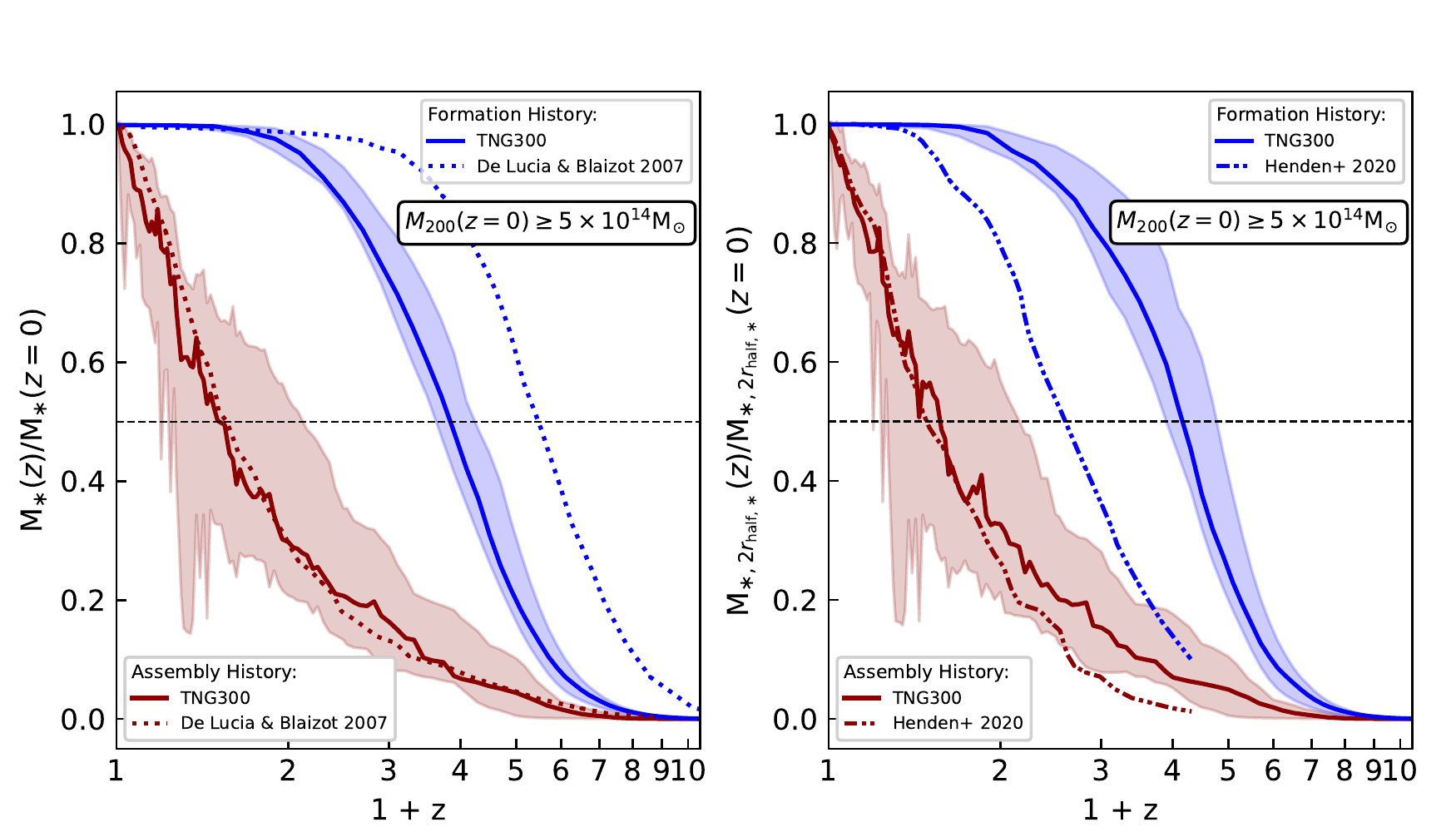}
	\caption{Median `assembly' (stars contained in the main progenitor, shown in red) and `formation' (all stars formed by a given time, shown in blue) histories of BCGs selected at $z=0$ within two different mass ranges: $\Mtwo(z=0) \geq 10^{14} \, \Msun$ (top panels) and $\Mtwo(z=0) \geq 5 \times 10^{14} \, \Msun$ (bottom panels). The left-hand panels compute the assembly and formation histories for the total stellar mass of the central subhalo ($M_{\ast}$), while the right-hand panels consider the stellar mass within twice the stellar half-mass radius ($M_{\ast, 2 r_{\rm half, \ast}}$). In both cases the stellar mass histories are normalized with respect to their value at $z=0$. The solid and dashed lines show results for the TNG300 and original Illustris simulations, respectively, while the dot-dashed line corresponds to the FABLE simulations \citep{Henden2020} and the dotted lines to the semi-analytic model of \protect\cite{DeLucia2007}. The shaded regions indicate the 16th to 84th percentile ranges of the TNG300 trends. For reference, the intersection between each stellar mass history and the horizontal dashed line indicates when the corresponding quantity reached half of its final value at $z=0$, i.e. the formation and assembly times.
	}
	\label{fig:SMF-vs-z}
\end{figure*}

\subsection{Formation versus assembly histories}
\label{subsec:formation_vs_assembly_histories}

Fig.~\ref{fig:SMF-vs-z} represents a different approach to studying the stellar mass growth of BCGs in the TNG300 simulation. Following \cite{DeLucia2007}, we compute and distinguish between the `formation' and `assembly' histories for our BCG sample. 

The assembly history (red lines) quantifies the stellar mass evolution of the main progenitor along the main branch in the merger trees, while the formation history (blue lines) represents the mass of all the stellar particles in the BCG at $z = 0$ that had already formed by a given time. Both stellar mass histories are normalized with respect to the stellar mass at $z=0$. These quantities are calculated for the total stellar mass of the main central galaxy (left-hand panels) and for the stellar mass of the main central galaxy within within 2$r_{\rm half, \ast}$ (right-hand panels). The solid lines show results for TNG300, while the shaded regions indicate the corresponding 16th to 84th percentile ranges. In the upper panels, which show haloes with masses at $z=0$ larger than $\Mtwo = 10^{14} \, \Msun$, the dashed lines show results from the original Illustris simulation. In the bottom panels, which show more massive haloes with $\Mtwo(z=0) \geq 5 \times 10^{14} \, \Msun$, the dotted lines show predictions from the semi-analytic model by \citet[][left]{DeLucia2007}, while the dot-dashed lines correspond to the FABLE simulations \citep[][right]{Henden2020} for $\Mtwo \geq 10^{14} \, \Msun$\footnote{By construction, the mass function of FABLE haloes is approximately flat (over logarithmic intervals), and the median halo mass of their $\Mtwo \geq 10^{14} \, \Msun$ sample is comparable to the median halo mass of our $\Mtwo \geq 5 \times 10^{14} \, \Msun$ sample.}. For reference, the intersections of the formation and assembly histories with the horizontal dashed line at a value of $0.5$ indicate the formation and assembly times, respectively.

We can see from Fig.~\ref{fig:SMF-vs-z} that, in TNG300, the stellar mass that eventually ends up in a typical BCG at $z=0$ (formation history, solid blue line) forms at early epochs, with half of it already created by $z \approx 2.7$--$3.2$ (with slight variations depending on the aperture and halo mass). However, this stellar mass did not assemble onto the main progenitor branch of the BCG (assembly history, solid red line) until later times, with the main progenitor reaching half of its final mass at $z \approx 0.8$ for $\Mtwo(z=0) \geq 10^{14} \, \Msun$ haloes, and at $z \approx 0.5$ for $\Mtwo(z=0) \geq 5 \times 10^{14} \, \Msun$ haloes. The original Illustris simulation (not shown in the bottom panels due to the absence of $\Mtwo \geq 5 \times 10^{14} \, \Msun$ objects within its smaller volume) displays very similar results to TNG300 for both stellar mass definitions, either using total stellar masses (left-hand panels) or stellar masses within 2$r_{\rm half, \ast}$ (right-hand panels).

Interestingly, the bottom panels of Fig.~\ref{fig:SMF-vs-z} show that the massive BCGs from the \cite{DeLucia2007} and FABLE \citep{Henden2020} galaxy formation models also assemble half of their final stellar mass around $z \approx 0.5$, just like in TNG300, suggesting that this result is a robust prediction of hierarchical cosmological models that is largely insensitive to differences in the implementation of galaxy formation physics.

We can also see from Fig.~\ref{fig:SMF-vs-z}, bottom-left panel, that the model of \cite{DeLucia2007} predicted that half of the stars that eventually become part of BCGs formed by $z \sim 4$--$5$, indicating instead that, in the case of SF times, the implementation of the feedback mechanisms does matter: the AGN feedback model in the semi-analytic model of \cite{DeLucia2007} is presumably more effective at even earlier times than that of Illustris or IllustrisTNG. On the other hand, the BCGs simulated by \cite{Henden2020} show `delayed' formation histories relative to TNG300, which is indicative of weaker AGN feedback.

Finally, \cite{ragone2018bcg}, using yet another set of zoom-in simulations of massive haloes with different feedback models, also computed the assembly and formation histories of simulated BCGs within different apertures, although their sample consisted of more massive objects ($\Mtwo \gtrsim 1.1 \times 10^{15} \, \Msun$). When considering the stellar mass within 10\% of $R_{500}$ (which is comparable to $2 r_{\rm half, \ast}$; see Appendix \ref{sec: appendix}), they obtained median assembly and formation redshifts of $z \approx 0.6$ and $z \approx 2.9$, respectively, in rough agreement with our TNG300 results.


\section{Stellar mass profiles around TNG300 BCGs}\label{sec:stellar_profiles}

\begin{figure*}
  \centering
	\includegraphics[width=15cm]{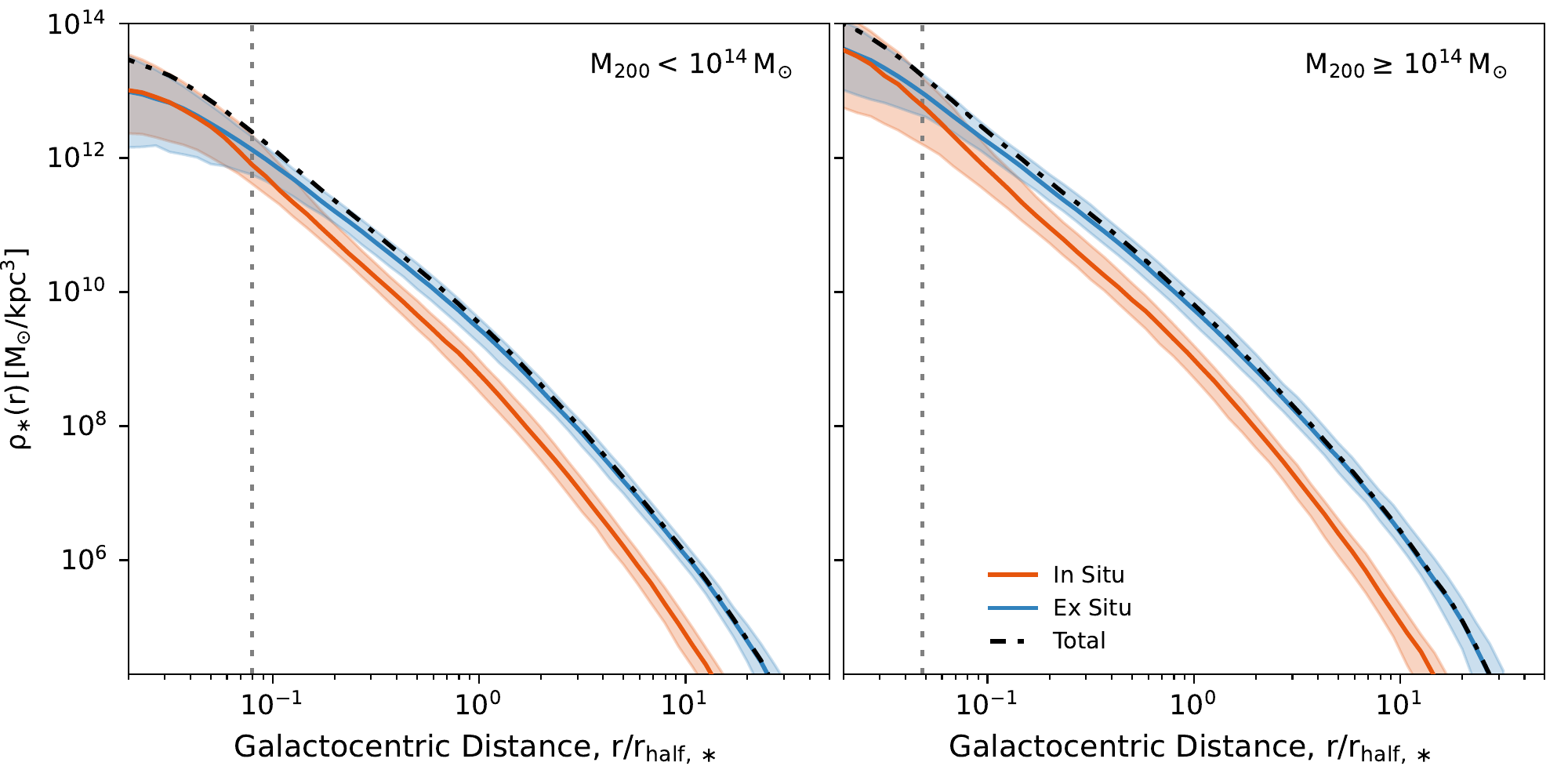}
	\includegraphics[width=15cm]{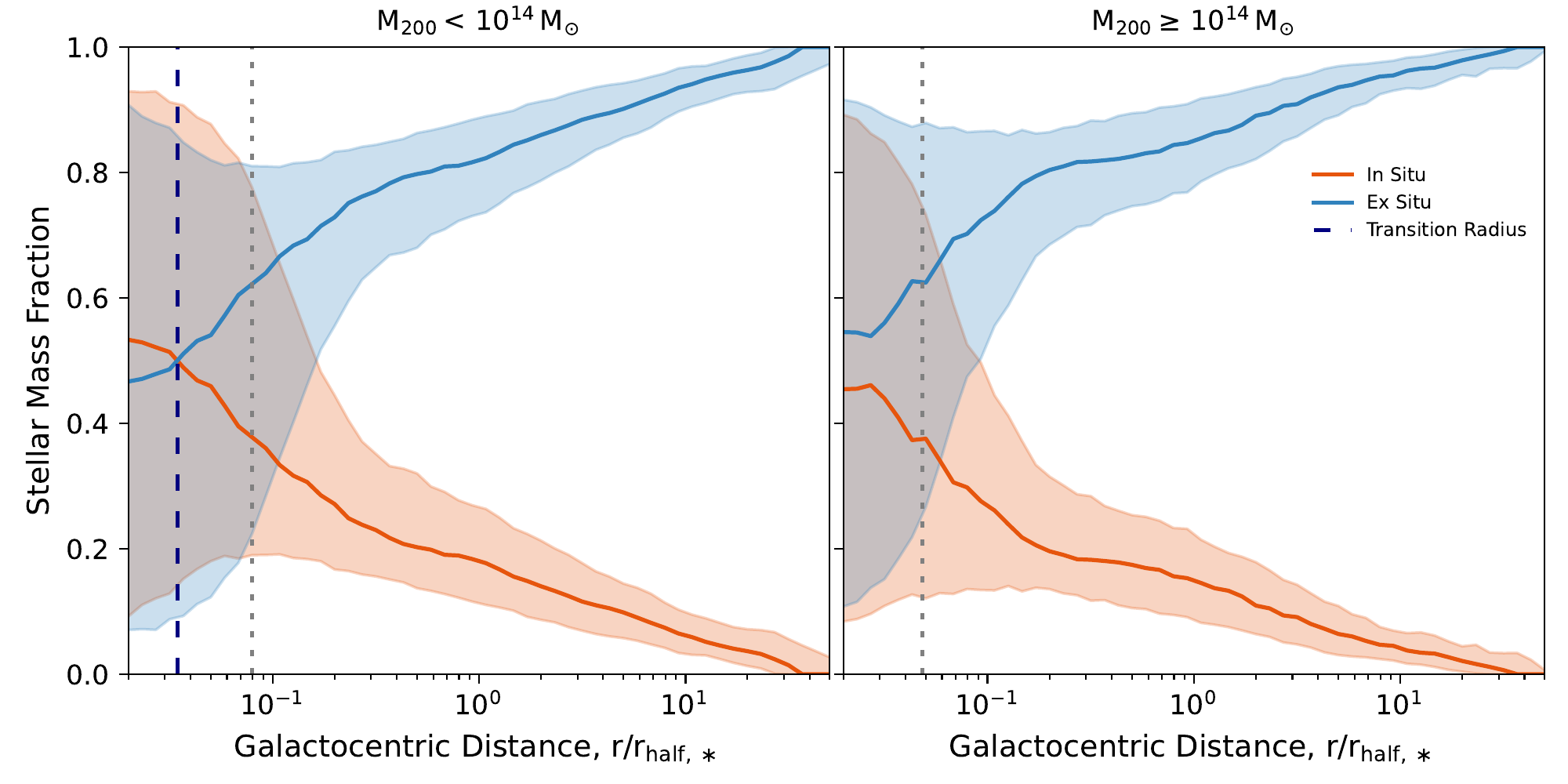}
	\caption{Stellar mass density profile and stellar mass fraction of BCGs in the TNG300 simulation. Upper panels: Median stellar mass density profiles (averaged over spherical shells) of TNG300 BCGs at $z = 0$ versus galactocentric distance (in units of the stellar half-mass radius, $r_{\rm half, \ast}$). Bottom panels: Median stellar mass fractions to total stellar mass (averaged over spherical shells) versus galactocentric distance (in units of the stellar half-mass radius, $r_{\rm half, \ast}$), shown for BCGs at $z=0$.
	The left-hand panels provide results for TNG300 haloes with $\Mtwo = 0.5$--$1 \times 10^{14} \, \Msun$, while the right-hand panels show $\Mtwo \geq 10^{14} \, \Msun$ objects. The black dot-dashed lines denote the total stellar mass density profiles, whereas the solid lines give the density of stars that were formed $in\ situ$ (orange) and $ex\ situ$ (blue). The shaded regions represent the (galaxy-to-galaxy) 16th to 84th percentile ranges of the latter two components, while the dashed grey line indicates where the limited numerical resolution suggests caution, shown here at 4 times the gravitational softening length at $z = 0$. In the bottom panels, the dashed navy blue line shows the \textit{transition radius}, which corresponds to the intersection between the $in\ situ$ and $ex\ situ$ components. Note that the transition radius falls below our adopted resolution limit for the less massive BCGs (left), while it is typically undefined for the more massive BCGs (right), since their profiles are usually dominated by accreted stars at all radii.}
	\label{fig:rho-vs-r&Stellar-mass-fraction-vs-r}
\end{figure*}

We conclude the analysis of the stellar mass content of BCGs in TNG300 by quantifying the spatial distribution at $z=0$ of the different stellar mass components of BCGs and their surrounding ICL by computing their stellar mass profiles. These profiles quantify how the stars in the BCG are distributed, at least according to the TNG300 simulation, as a function of the distance to the centre of the BCG. Throughout this section we again use the \textit{in situ} and \textit{ex situ} stellar classifications.

The upper panels of Fig.~\ref{fig:rho-vs-r&Stellar-mass-fraction-vs-r} show the stellar mass density of TNG300 BCGs at $z=0$ as a function of the galactocentric distance, $r$, normalized by the stellar half-mass radius, $r_{\rm half, \ast}$. The left-hand and right-hand panels show the median density profiles for BCGs (including the ICL component) with halo masses $\Mtwo = 0.5$--$1 \times 10^{14} \, \Msun$ and $\Mtwo \geq 10^{14} \, \Msun$, respectively, averaged over spherical shells. The orange and blue solid lines show the density of \textit{in situ} and \textit{ex situ} stars, respectively, while the black dot-dashed line corresponds to the total gravitationally bound stellar component. The vertical dotted grey line indicates the spatial resolution limit of the simulation, which here we define as four times the gravitational softening length in TNG300 at $z=0$, $4\epsilon \approx 5.9 \, {\rm kpc}$ (see \citealt{Pillepich2018, Pillepich2018a} for additional comments on the effects of numerical resolution on galaxies' stellar masses and their spatial distribution).

The upper panels of Fig.~\ref{fig:rho-vs-r&Stellar-mass-fraction-vs-r} indicate that, according to TNG300, \textit{ex situ} stars are the dominant component in BCGs, especially at large distances from the galactic centre, where the ICL component dominates, and that the slope of the \textit{ex situ} stellar mass density profile is shallower than that of the \textit{in situ} component. In less massive galaxies, the galactic centre is typically dominated by \textit{in situ} stars, which implies that there exists a \textit{transition radius} \citep{pillepich2015building, Rodriguez-Gomez2016} where \textit{in situ} and \textit{ex situ} stars become equally abundant. Importantly, however, the upper panels of Fig.~\ref{fig:rho-vs-r&Stellar-mass-fraction-vs-r} demonstrate  that objects as extreme as the most massive BCGs can be dominated by \textit{ex situ} stars at essentially all radii. With this, we confirm and further emphasize analogous results pointed out by \cite{Pillepich2018a} and \cite{pulsoni2021}.

\begin{figure*}
  \centering
  \vbox{
	\includegraphics[width=15cm]{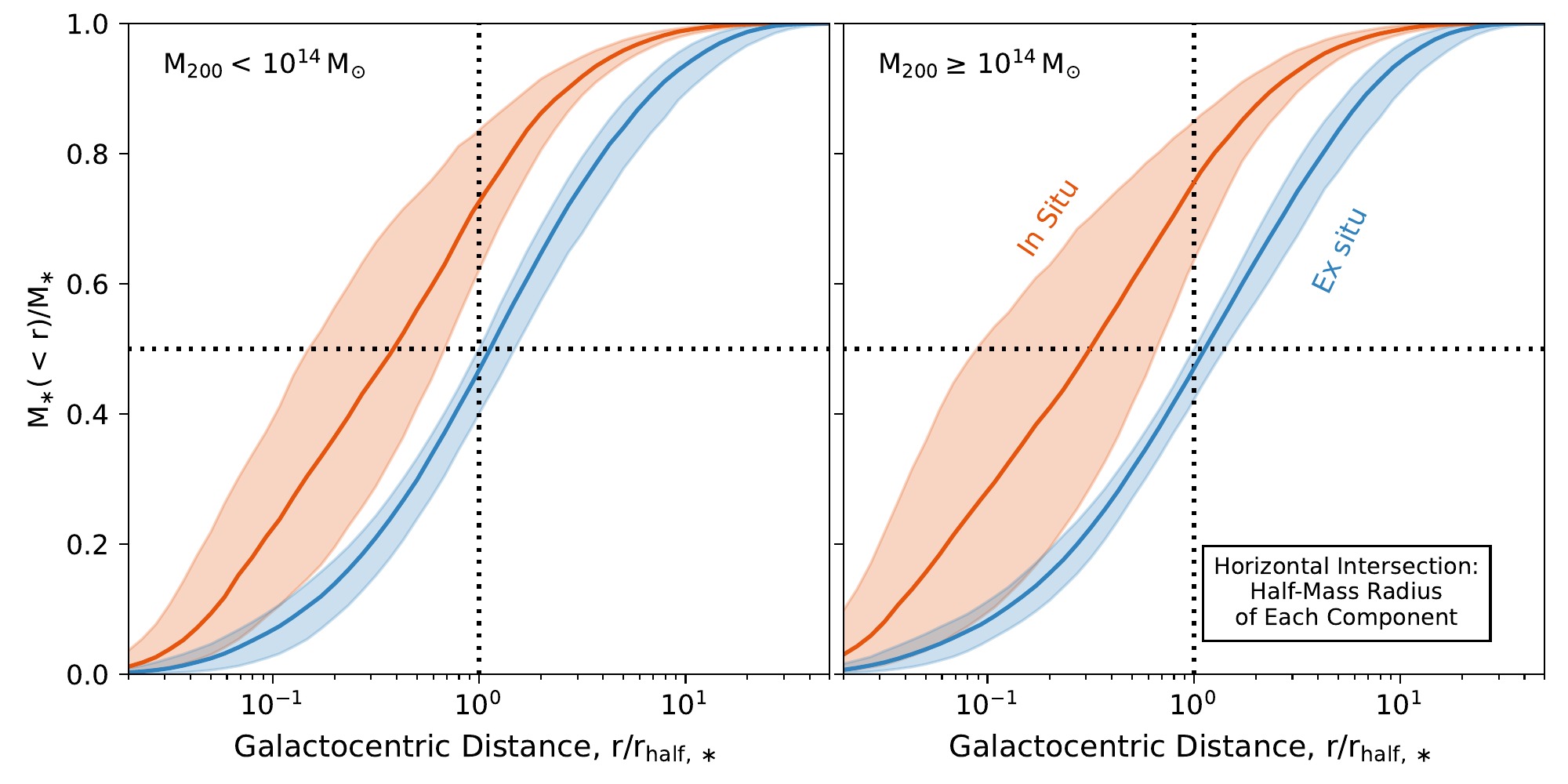}
	\\
	\includegraphics[width=15cm]{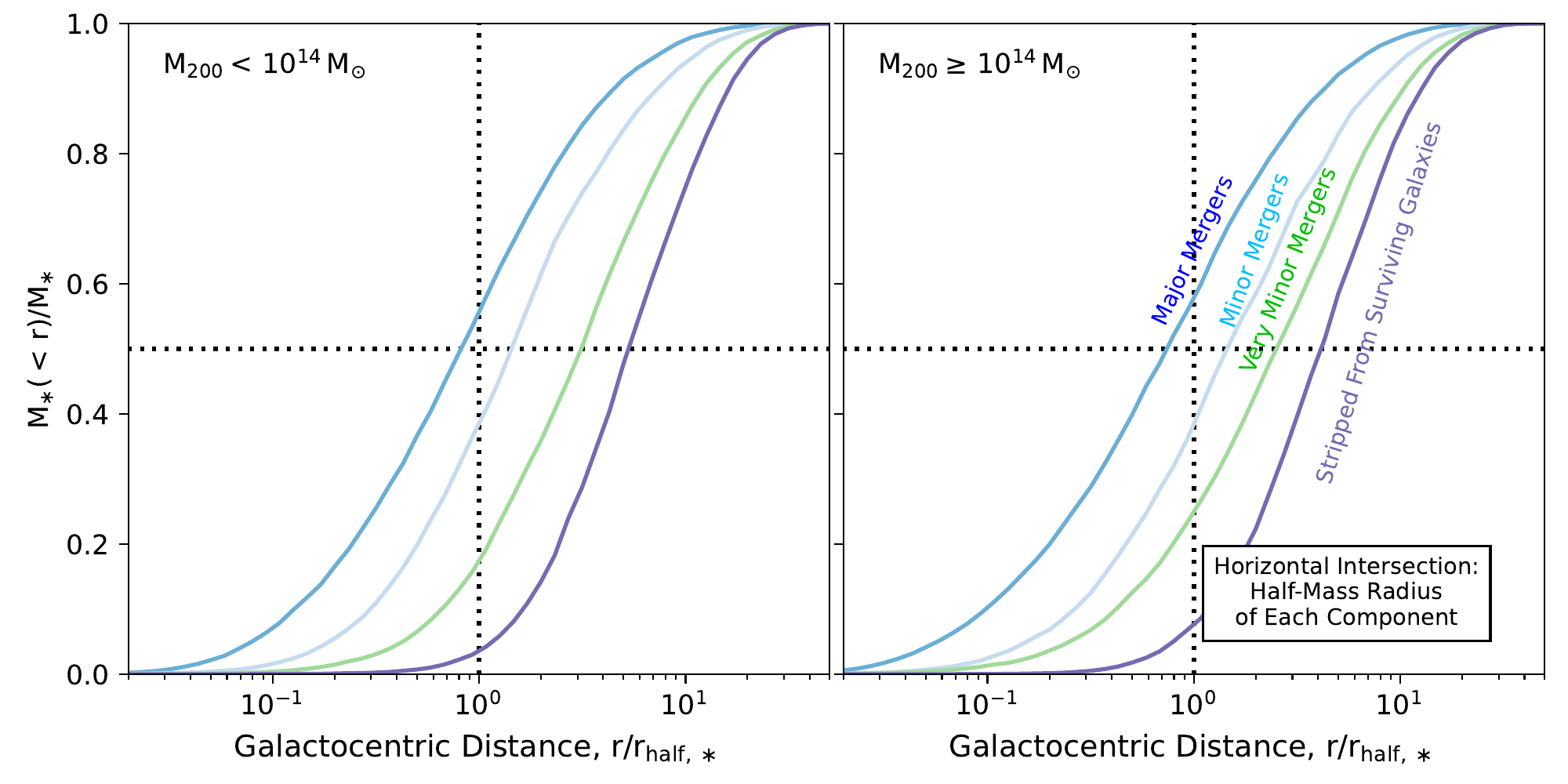}
	}
	\caption{Median cumulative mass profiles of the different stellar components of BCGs from the TNG300 simulation at $z = 0$ as a function of galactocentric distance (normalized by the stellar half-mass radius, $r_{\rm half, \ast}$), shown for different halo mass ranges: $\Mtwo = 0.5$--$1 \times 10^{14} \, \Msun$ (left) and $\Mtwo \geq 10^{14} \, \Msun$ (right). The upper panels show the cumulative distribution of \textit{in situ} and \textit{ex situ} stars, while the lower panels split the \textit{ex situ} component into stars that were brought in by mergers of different mass ratios and stars that were stripped from surviving galaxies. The profiles are normalized with respect to the total mass of each component. The shaded regions show the 16th to 84th percentile ranges. For reference, the vertical and horizontal dotted lines indicate the stellar half-mass radius ($r_{\rm half, \ast}$) and half of the mass in each component, respectively.}
	\label{fig:comulative_rho-vs-r}
\end{figure*}

In the bottom panels of Fig.~\ref{fig:rho-vs-r&Stellar-mass-fraction-vs-r}, we compute the median stellar mass fraction (with respect to total stellar mass) of the \textit{in situ} and \textit{ex situ} components within spherical shells, again as a function of galactocentric distance normalized by the stellar half-mass radius, $r / r_{\rm half, \star}$, for the same halo mass bins (different panels). This figure exhibits more clearly where the transition radius occurs, at least for the median profiles. For the lower halo mass range ($\Mtwo = 0.5$--$1 \times 10^{14} \, \Msun$, left-hand panel), there is a transition radius but it is typically smaller than the spatial resolution limit of the simulation, and may not be properly captured. For the higher halo mass range ($\Mtwo \geq 10^{14} \, \Msun$, right-hand panel), according to TNG300, the \textit{in situ} and \textit{ex situ} profiles typically do not intersect, indicating that accreted stars dominate the stellar mass content down to the galactic centre. This contrasts with results by \cite{Rodriguez-Gomez2016} for the original Illustris simulation, for which it was possible to define a transition radius for the vast majority of galaxies with stellar masses $M_{\ast} = 10^{9} - 10^{12} \, \Msun$. This is a consequence of the smaller size of the original Illustris volume (106.5 Mpc per side), which produced a smaller number of BCGs, combined with less effective AGN feedback in massive galaxies, which resulted in a higher amount of \textit{in situ} stars.

Using the TNG100 simulation, \cite{pulsoni2021} found that some of their early-type galaxies (with stellar masses $M_{\ast} \geq 10^{10.3} \, \Msun$) were dominated by accreted stars at all radii (namely, their `class 4' galaxies). However, these objects only constituted $\sim$8 per cent of their galaxy sample, with the vast majority showing more `classic' stellar mass profiles, dominated by \textit{in situ} stars in their inner regions. Similarly, using the Magneticum cosmological simulation \citep{hirschmann2014}, \cite{remus2021accreted} studied the stellar density profiles of $\sim$500 galaxies with $M_{\ast} \geq 2 \times 10^{10} \, \Msun$, finding that $\sim$9 per cent of them were dominated by accreted stars at all radii. We note that these previous works considered smaller cosmological volumes (110.7 and 68 Mpc per side for the TNG100 and Magneticum simulations, respectively), so they did not include many massive BCGs in their sample, by construction. Fig.~ \ref{fig:rho-vs-r&Stellar-mass-fraction-vs-r} shows that, for our highest-mass BCG sample in TNG300, the absence of a transition radius (or a transition radius below the resolution limit) is the norm rather than the exception.

Finally, in Fig.~\ref{fig:comulative_rho-vs-r} we compute the median of the cumulative stellar mass profiles as a function of galactocentric radius, normalized by $r_{\rm half, \star}$, for the same halo mass bins considered in Figs \ref{fig:rho-vs-r&Stellar-mass-fraction-vs-r}. The upper panels show the cumulative profiles of the \textit{in situ} (orange) and \textit{ex situ} (blue) stellar components, with shaded regions indicating the 16th to 84th percentile ranges, while the lower panels split the \textit{ex situ} component into stars that were accreted via major mergers ($\mu \geq 1/4$), minor mergers ($1/10 \leq \mu < 1/4$), very minor mergers ($\mu < 1/10$), and stars that were stripped from surviving galaxies. Each profile is normalized by the total mass of its respective component, so that its intersection with the horizontal dotted line (at a value of 0.5) indicates the radius that encloses half of the mass of the respective component.

Fig.~\ref{fig:comulative_rho-vs-r} shows that \textit{in situ} stars have a more concentrated spatial distribution than \textit{ex situ} stars, while the latter display a spatial segregation based on merger mass ratio: stars accreted in major mergers are found closer to the centre, followed by minor mergers, very minor mergers, and finally stars that were stripped from surviving galaxies. These results are in qualitative agreement with those found by \cite{Rodriguez-Gomez2016} for the original Illustris simulation, as well as with theoretical expectations  \citep[e.g.][]{amorisco2017}. BCGs are located in very deep gravitational potential wells, such that it is difficult to remove stars that were born \textit{in situ}, while \textit{ex situ} stars arrive from other galaxies with high amounts of angular momentum and kinetic energy, which need to be removed through some physical mechanism in order for them to reach the centre of the BCG. One such mechanism for losing angular momentum is dynamical friction, which primarily affects the orbits of more massive satellite galaxies \citep{chandrasekhar1943dynamical}, leading to the spatial segregation of accreted stars by merger mass ratio seen in Fig.~\ref{fig:comulative_rho-vs-r}.


\section{Summary and Discussion}
\label{sec:discussion_and_conclusions}

We have used the TNG300 simulation of the IllustrisTNG project to study the formation and evolution of different stellar components in galaxy clusters, focusing on the massive galaxies at their centres: the BCGs. Our cluster sample at $z=0$ is defined by all haloes with $\Mtwo \geq 5 \times 10^{13} \, \Msun$, which in TNG300 constitutes a total of 700 systems (including three haloes with $\Mtwo \geq 10^{15}\, \Msun$) -- for comparison, this selection would yield 41 objects in the smaller TNG100 simulation. For each halo and throughout the paper, we have considered the BCG and ICL components to be composed, by construction, of stars that are gravitationally bound to the main central galaxy, identified by \textsc{subfind} as the one occupying the minimum of the halo potential well (see Table~\ref{tab: Definition components} for the operational definitions adopted throughout). Within this framework, we have expanded upon the analysis of \cite{Pillepich2018a} and compared the stellar mass content of our clusters with observations and other simulations, investigated the origin of the stars in the BCG and ICL, explored the redshift evolution of the SFRs and stellar masses of BCGs, and characterized the spatial distribution of the different stellar mass components in BCGs at $z=0$. Throughout this section, we summarize and discuss the main results of this work.

{\bf Stellar masses versus observations.} By computing the stellar masses of our simulated BCG+ICL systems within a fixed (3D) 100 kpc aperture and plotting them as a function of halo mass ($M_{500}$), we find broad agreement with observations by \cite{DeMaio2020} at different redshifts, from $z \approx 0.1$ to $z \approx 0.4$ (Fig.~\ref{fig:Mstellar-Mhalo in z}). However, the comparison at $z \approx 1.5$ is inconclusive, since there is little overlap between the halo masses of simulated and observed clusters at such redshifts. More detailed comparisons to observations will be carried out in upcoming work.

{\bf Major and minor mergers.}
Using the simulation merger trees, we have shown that the TNG300 BCGs have experienced, on average, $\sim$2, $\sim$3, and $\sim$5 major mergers (stellar mass ratio $\mu \geq 1/4$) since $z=1$, $z=2$, and in their entire histories, respectively (Fig.~\ref{fig: number-mergers-and-ex-situ_star-fraction}, upper panel). The number of major+minor mergers ($\mu \geq 1/10$) is $\sim$1.5--2 times larger than the number of major mergers (Fig.~\ref{fig: number-mergers-and-ex-situ_star-fraction}, middle panel). The fraction of stellar mass contributed by \textit{ex situ} stars -- i.e. those that were formed in other galaxies and subsequently accreted -- is approximately 70, 80, and 90 per cent for the BCG, BCG+ICL, and ICL, respectively, with little dependence on halo mass up to $\Mfive \approx 2 \times 10^{14} \, \Msun$ (Fig.~\ref{fig: number-mergers-and-ex-situ_star-fraction}, bottom panel). These fractions increase by $\sim$5--10 per cent for the most massive systems in the simulation, with halo masses approaching $\Mfive \sim 10^{15} \, \Msun$.

{\bf Stellar mass composition of BCGs and ICL.}
In order to quantify the origin of the stellar mass in the BCG and ICL in more detail, we have created pie charts that present not only the average fraction of stellar mass contributed by \textit{in situ} and \textit{ex situ} stars, but also how the latter component splits into stars that were accreted in major ($\mu \geq 1/4$), minor ($1/10 \leq \mu < 1/4$), and very minor ($\mu < 1/10$) mergers, as well as stars that were stripped from surviving galaxies (Fig.~\ref{fig:Pie-chart}). For both halo mass ranges ($\Mtwo = 0.5$--$1 \times 10^{14} \, \Msun$ and $\Mtwo \geq 10^{14} \, \Msun$), we again find that the BCG, BCG+ICL, and ICL have overall \textit{ex situ} stellar mass fractions of approximately 70, 80, and 90 per cent, respectively. Additionally, we show that, according to TNG300, the fraction of the total stellar mass contributed by \textit{ex situ} stars that were accreted in major mergers is higher in the BCG ($\sim$40 per cent) than in the ICL ($\sim$25--30 per cent), while the contribution from \textit{ex situ} stars that were tidally stripped from surviving galaxies is much higher in the ICL ($\sim$30--40 per cent) than in the BCG ($\sim$2 per cent). Finally, the stellar mass fractions from stars that were accreted in minor and very minor mergers are similar in the BCG and ICL, ranging between 10 and 15 per cent.

{\bf SFRs of BCGs versus observations at different redshifts.}
We have calculated the median SFR and sSFR of BCGs in clusters more massive than $\Mtwo = 10^{14}$ M$_\odot$ at different redshifts, and compared them to several observational inferences at $z \lesssim 2$ (Fig.~\ref{fig:SFR-vs-z_observation}). We find that the SFRs and sSFRs of our simulated BCGs are in good agreement with observations at low redshift ($z \lesssim 0.4$) by \cite{Orellana-Gonzalez+2022} after imposing an SFR threshold of $0.1 \, \Msun \, {\rm yr^{-1}}$, while higher-redshift observations by \cite{mcdonald2016SFR} and \cite{Bonaventura2017} display higher values of SFR and sSFR than the TNG300 simulation, especially at $z \gtrsim 1$. As discussed in Section \ref{subsec:sfr_evolution}, several selection effects, observational limitations, and model uncertainties affect current observational inferences, mainly in the direction of overestimating the SFRs and sSFRs plotted in Fig.~\ref{fig:SFR-vs-z_observation}. Thus, taking into account such overestimates, we cannot exclude that results from TNG300 could in fact be consistent with SFR observations at $z \gtrsim 1$, although more detailed comparisons need to be performed.

{\bf SF histories and mass assembly of BCGs.}
By tracking the TNG300 BCGs selected at $z=0$ back in time along their main progenitor branches in the merger trees, we calculated their median SFR and sSFR histories. The SFR histories show that the BCG progenitors quickly increased their SFRs at very early epochs, reaching a maximum at $z \sim 3$--$4$ and then decreasing strongly around $z \sim 2$--$3$ (Fig.~\ref{fig:SFR-vs-z_simulation}, top). Similarly, the sSFR histories show that the BCG progenitors were star-forming galaxies at $z \gtrsim 3$, but by $z \sim 1$ most of them had already transitioned to a quiescent regime (Fig.~\ref{fig:SFR-vs-z_simulation}, bottom). In the IllustrisTNG model, this is due to the effects of supermassive BH feedback. Furthermore, the SFR and sSFR histories of our BCGs, as well as their stellar, halo, and central BH mass histories (Fig. \ref{fig:mass-vs-z}), are in good agreement with semi-empirical inferences by Rodr\'iguez-Puebla et al. (in preparation), this comparison not being subject to potential biases due to selection effects.

The median stellar mass history of TNG300 BCGs shown in Fig.~\ref{fig:mass-vs-z} displays a bend around $z \sim 3$--$4$, which is directly related to the fact that the \textit{in situ} stellar mass growth slows down significantly in this redshift range because of quenching. On the other hand, the \textit{ex situ} stellar mass, which is comparatively small at early times, does not present such a bend and keeps growing until it eventually surpasses the \textit{in situ} component at $z \sim $1--2 (Fig.~\ref{fig:mstar(in-ex)situ-vs-z}). This is a vivid illustration of the `two-phase' formation process of massive galaxies \citep[e.g.][]{Oser2010, Rodriguez-Gomez2016}, which according to TNG300 also provides a good description of BCG stellar mass assembly.

In order to gain additional insight into the stellar mass growth of BCGs, we have also computed the median `assembly' and `formation' histories previously studied by \cite{DeLucia2007}, which quantify the stellar mass history of the main progenitor of the BCG (assembly history) and the stellar mass contributed by all the stars formed before a given time that are ultimately part of the BCG at $z= 0$ (formation history). Applying this formalism to the TNG300 simulation, we find that half of the stars that ultimately end up in BCGs had formed by $z \sim 3$, mostly in other galaxies, while the BCGs themselves reached half of their final stellar mass by $z \approx 0.8$ for BCGs with $\Mtwo(z = 0) \geq 10^{14} \, \Msun$, and by $z \approx 0.5$ for BCGs with $\Mtwo(z = 0) \geq 5 \times 10^{14} \, \Msun$ (Fig.~\ref{fig:SMF-vs-z}). Interestingly, the latter result is in excellent agreement with theoretical predictions by \cite{DeLucia2007} and \cite{Henden2020}. Therefore, the original result from \citet{DeLucia2007} that massive BCGs reach half of their final stellar mass around $z \approx 0.5$ appears to be a very robust prediction of the $\Lambda$CDM cosmological framework, being essentially independent of the galaxy formation model and numerical technique (hydrodynamic cosmological simulations versus semi-analytic models).

On the other hand, the formation time of the stars that ultimately end up in a BCG is more model-dependent. Half of the stars that ultimately constitute a massive BCG at $z=0$ were formed earlier ($z \approx 4$--$5$) in the model of \cite{DeLucia2007}, which is indicative of a possibly stronger AGN feedback than in IllustrisTNG, while the BCGs in the FABLE simulations \citep{Henden2020} show somewhat `delayed' formation histories compared to IllustrisTNG. Ultimately, these differences are perhaps not surprising given that these different models make different predictions for other observables, e.g. different quenched fractions of massive galaxies at different epochs, which may be more or less consistent with observations depending on the model \citep[see][for a comparison of quenched fractions across current cosmological galaxy simulations]{donnari2021b}.

{\bf Stellar mass profiles of the different components.}
Finally, we studied the stellar mass profiles of TNG300 BCGs at $z=0$. By calculating their stellar mass density in spherical shells, we have shown that the \textit{ex situ} stellar component dominates the outer regions and has a shallower slope than the \textit{in situ} component (Fig.~\ref{fig:rho-vs-r&Stellar-mass-fraction-vs-r}, upper panels). In more detail, by plotting the fractions of \textit{in situ} and \textit{ex situ} stellar mass as a function of radius, we showed more clearly that the stellar mass profiles of BCGs are often dominated by \textit{ex situ} stars down to the galactic centre, such that a `transition radius', where \textit{in situ} and \textit{ex situ} stars become equally abundant \citep[e.g.][]{pillepich2015building, Rodriguez-Gomez2016, pulsoni2021}, cannot be defined robustly for most BCGs (Fig.~\ref{fig:rho-vs-r&Stellar-mass-fraction-vs-r}, bottom panels). By plotting the cumulative mass profiles of different stellar components, we also demonstrated that the \textit{ex situ} stars are spatially distributed according to merger mass ratio: stars that were accreted in major mergers tend to be found closer to the centre, followed by those from minor mergers, very minor mergers, and finally stars that were stripped from surviving galaxies (Fig.~\ref{fig:comulative_rho-vs-r}). These results show qualitative agreement with a previous analysis for the Illustris simulation \citep{Rodriguez-Gomez2016}, as well as with expectations from simple theoretical arguments (e.g. dynamical friction).

\vspace{1em}

In summary, we have presented a theoretical study of the formation and evolution of BCGs using the TNG300 cosmological simulation, which, due to its large volume ($\sim$300 Mpc per side), is ideal for studying these rare objects in a cosmological context. We have found that, taken at face value, certain aspects of BCGs in TNG300 might be in tension with some of the observational studies we considered, but, as we discussed, a better agreement could be reached through forward-modelling of the different observables that we compare against. On the other hand, the TNG300 predictions agree well with the inferences from a sophisticated semi-empirical approach, which by construction is consistent with a large body of observed galaxy distributions and correlations from $z\sim0$ to $z\sim 10$. Our results show that the evolution of BCGs fits very well within the two-phase formation scenario of early-type galaxies \citep[e.g.][]{Oser2010, Rodriguez-Gomez2016}, with an early dissipative phase of vigorous \textit{in situ} SF ($z \gtrsim 3$--$4$) followed by an abrupt drop in SF at $z \sim 2$--$3$, and then a longer phase of non-dissipative stellar mass growth driven by major and minor mergers that contribute \textit{ex situ} stars, which constitute most of the stellar mass in the BCGs after $z \sim 1$--$2$. Differently from less extreme galaxies, our simulated BCGs are ultimately dominated by \textit{ex situ} stars at $z=0$, even in their central regions, which makes these galaxies very interesting. This is not surprising given that BCGs form at the centres of the gravitational potential wells of the most massive structures of the Universe, such that their evolution is driven by more extreme environmental mechanisms.

\section*{Acknowledgements}

We thank Bernardo Cervantes-Sodi for useful comments and discussions. We are also grateful to the referee, Emanuele Contini, for an insightful report that helped to improve the manuscript. VRG and LVS acknowledge support from UC MEXUS-CONACyT grant CN-19-154. ARP acknowledges support from the CONACyT `Ciencia Basica' grant 285721. The IllustrisTNG flagship simulations were run on the HazelHen Cray XC40 supercomputer at the High Performance Computing Center Stuttgart (HLRS) as part of project GCS-ILLU of the Gauss Centre for Supercomputing (GCS). Ancillary and test runs of the project were also run on the compute cluster operated by HITS, on the Stampede supercomputer at TACC/XSEDE (allocation AST140063), at the Hydra and Draco supercomputers at the Max Planck Computing and Data Facility, and on the MIT/Harvard computing facilities supported by FAS and MIT MKI. The Flatiron Institute is supported by the Simons Foundation.

\section*{Data availability}

The data from the Illustris and IllustrisTNG simulations used in this work are publicly available at the websites \href{https://www.illustris-project.org}{https://www.illustris-project.org} and \href{https://www.tng-project.org}{https://www.tng-project.org}, respectively \citep{Nelson2015, nelson2019illustristng}.

\bibliography{paper}
\bibliographystyle{mnras}

\appendix

\section{Comparison of stellar masses measured within different apertures}
\label{sec: appendix}

\begin{figure*}
  \centering
	\includegraphics[width=14.2cm]{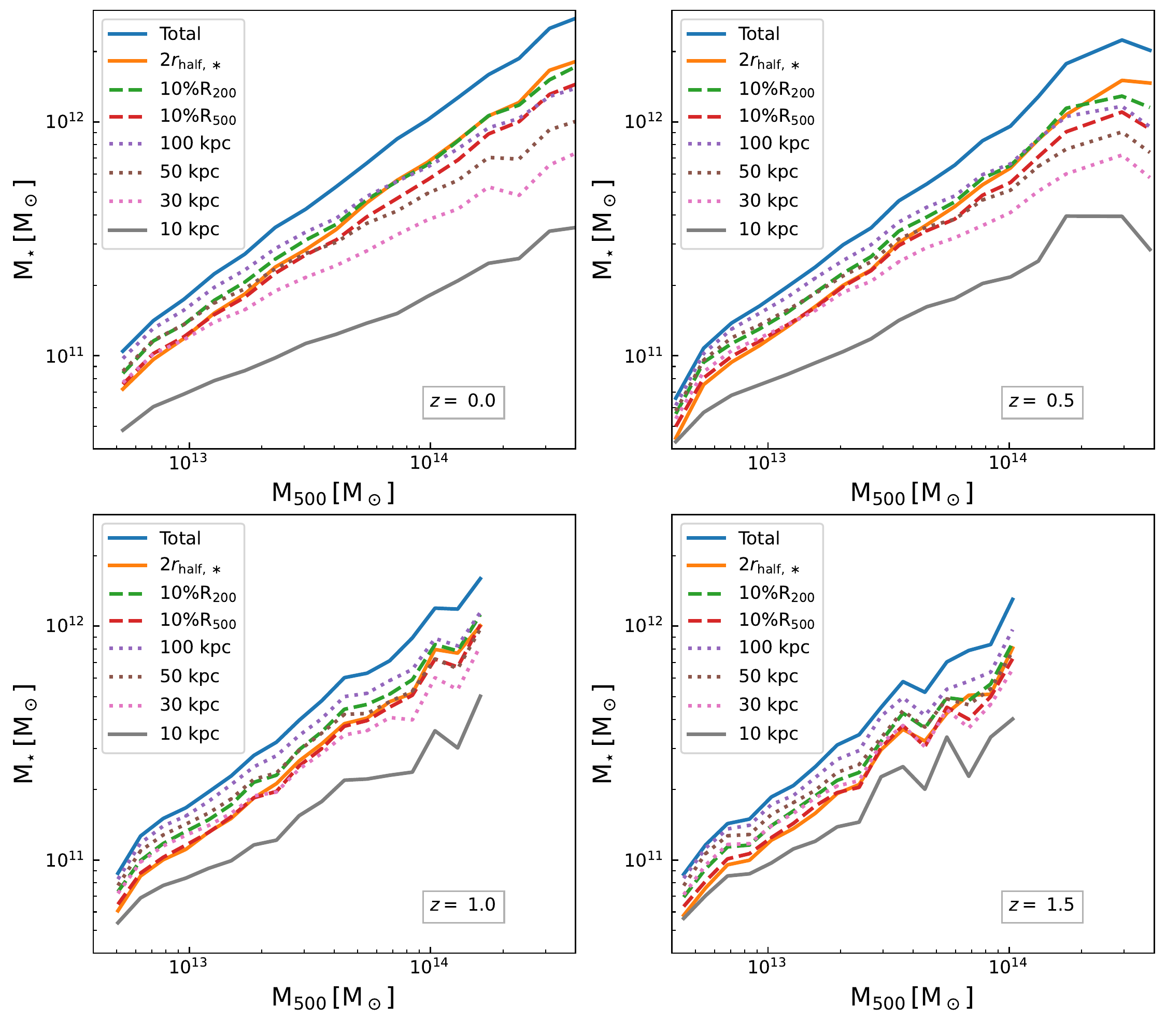}
	\caption{Median stellar mass as a function of halo mass ($\Mfive$) at different redshifts (different panels). The different lines correspond to the stellar mass measured within different spherical apertures: for fixed radii in physical units (10, 30, 50, and 100 kpc) and for apertures that scale with the size of the galaxy or halo ($2 r_{\rm half, \ast}$, $0.1 \Rtwo$, and $0.1 R_{500}$), as indicated in the legend. The solid blue line shows the total stellar mass of the central subhalo.}
	\label{fig:Stellar-mass-halo-mass-diff-apertura}
\end{figure*}

Fig.~\ref{fig:Stellar-mass-halo-mass-diff-apertura} shows the median stellar mass as a function of halo mass ($\Mfive$) for different aperture definitions at redshifts ($z = 0.0, 0.5, 1.0 \, \rm{and} \, 1.5$). The blue lines show the total stellar mass of the central \textsc{subfind} objects, while the orange lines show the stellar mass within twice the stellar half-mass radius, $r_{\rm half, \ast}$. The dashed green and red lines correspond to the stellar mass within 10 per cent of $\Rtwo$ and $R_{500}$, respectively. The dotted lines show the stellar mass computed within fixed (3D) apertures of 30, 50, and 100 kpc, while the solid grey line represents the stellar mass within a fixed aperture of 10 kpc.

\end{document}